\documentclass[%
reprint,
superscriptaddress,
 amsmath,amssymb,
 aps,
prd,
]{revtex4-2}

\usepackage[utf8]{inputenc}
\usepackage{graphicx}
\usepackage{dcolumn}
\usepackage{multirow}
\usepackage{bm}
\usepackage{mathrsfs}
\usepackage{hyperref}
\hypersetup{
  colorlinks = true,
  urlcolor = blue,
  linkcolor = blue,
  citecolor = green,
  filecolor = magenta,
}

\newcommand{\ud}{\mathrm{d}}
\newcommand{\pd}{\partial}

\newcommand{\lie}{\mathscr{L}}
\newcommand{\sinv}[2]{s_{#1}^{-#2}}

\newcommand{\order}[1]{\mathcal{O}\left(#1\right)}

\newcommand{\cA}{\mathcal{A}}

\newcommand{\tdu}{\tilde{u}}
\newcommand{\tdr}{\tilde{r}}

\newcommand{\tdn}{\tilde{n}}

\newcommand{\tde}{\tilde{e}}
\newcommand{\tdth}{\tilde{\theta}}
\newcommand{\tpd}{\tilde{\partial}}
\newcommand{\tdcd}{\tilde{\mathscr D}}
\newcommand{\tdgamma}{\tilde{\gamma}}

\newcommand{\tdg}{\tilde{g}}

\newcommand{\mfu}{\mathfrak{u}}
\newcommand{\mfv}{\mathfrak{v}}
\newcommand{\tae}{\mathfrak{u}}
\newcommand{\mfa}{\mathfrak{a}}
\newcommand{\mfx}{\mathfrak{x}}
\newcommand{\mfy}{\mathfrak{y}}

\newcommand{\sfU}{\mathsf{U}}
\newcommand{\sfR}{\mathsf{R}}
\newcommand{\sfm}{\mathsf{m}}
\newcommand{\sfg}{\mathsf{g}}
\newcommand{\sfv}{\mathsf{v}}
\newcommand{\sfs}{\mathsf{s}}
\newcommand{\sfk}{\mathsf{k}}
\newcommand{\sfc}{\mathsf{C}}
\newcommand{\sfd}{\mathsf{D}}

\newcommand{\sD}{\mathscr{D}}
\newcommand{\sW}{\mathscr{W}}
\newcommand{\sV}{\mathscr{V}}
\newcommand{\sT}{\mathscr{T}}
\newcommand{\sR}{\mathscr{R}}

\newcommand{\ebms}{\overline{\text{BMS}}}

\begin{document}


\title{Asymptotic analysis of Einstein-\AE{}ther theory and its memory effects: the linearized case}

\author{Shaoqi Hou}
\email{hou.shaoqi@whu.edu.cn}
\affiliation{School of Physics and Technology, Wuhan University, Wuhan, Hubei 430072, China}
\author{Anzhong Wang}
\email{anzhong\_wang@baylor.edu}
\affiliation{GCAP-CASPER, Physics Department, Baylor University, Waco, Texas 76798-7316, USA}
\author{Zong-Hong Zhu}
\email{zhuzh@whu.edu.cn}
\affiliation{School of Physics and Technology, Wuhan University, Wuhan, Hubei 430072, China}
\affiliation{Department of Astronomy, Beijing Normal University, Beijing 100875,  China}

\date{\today}

\begin{abstract}
  This work analyzes the asymptotic behaviors of the asymptotically flat solutions of Einstein-\ae ther theory in the linear case. 
  The vacuum solutions for the tensor, vector, and scalar modes are first obtained, written as sums of various multipolar moments.  
  The suitable coordinate transformations are then determined, and the so-called pseudo-Newman-Unti coordinate systems are constructed for all radiative modes.
  In these coordinates, it is easy to identify the asymptotic symmetries. 
  It turns out that all three kinds of modes possess the familiar Bondi-Metzner-Sachs symmetries or the extensions as in general relativity.
  Moreover,  there also exist the \emph{subleading} asymptotic symmetries parameterized by a time-independent vector field on a unit 2-sphere.
  The memory effects are also identified. 
  The tensor gravitational wave also excites similar displacement, spin, and center-of-mass memories to those in general relativity.
  New memory effects due to the vector and scalar modes exist.
  The subleading asymptotic symmetry is related to the (leading) vector displacement memory effect, which can be viewed as a linear combination of the electric-type and magnetic-type memory effects. 
  However, the scalar memory effect seems to have nothing to do with the asymptotic symmetries at least in the linearized theory.
\end{abstract}

\maketitle


\section{Introduction}

The gravitational wave (GW) not only causes the relative distance of two adjacent test particles to oscillate, but also leaves a trace of its ever existence, after it disappears.  
This trace is nothing but the memory effect, the permanent change in the relative distance between test particles. 
It was first discussed theoretically as early as 1970s \cite{Zeldovich:1974gvh,Braginsky:1986ia}.
It was found that the permanent change is proportional to the total variation in the quadrupole moment of the source of gravity before and after the GW emission \cite{1987Natur.327..123B}. 
So this effect is also named the linear memory.
In 1990s, the nonlinear memory effect was uncovered, which is sourced by the GW itself, or any null radiation \cite{Christodoulou1991,Wiseman:1991ss,Blanchet:1992br,Thorne:1992sdb}.
In the literature, the linear memory is also said to be ordinary, while the nonlinear one is null \cite{Bieri:2013ada}.

Symmetries have played important roles in modern theoretical physics \cite{Wald:1984rg,Weinberg:1995mt,Srednicki:2007qs,Schwartz:2013pla}.
Although in a generic gravitating system, there are no spacetime symmetries, the asymptotically flat spacetime possesses the so-called Bondi-Metzner-Sachs (BMS) symmetry \cite{Sachs1962asgr}.
It is a kind of diffeomorphism that preserves the boundary conditions of the spacetime at the null infinity in the Bondi-Sachs coordinates \cite{Bondi:1962px,Sachs:1962wk}.
So under this coordinate transformation, the asymptotic behavior of the metric is unchanged.
The corresponding symmetry group, named BMS group, is a semi-direct product of the Lorentz group by the supertranslation group. 
The supertranslation generalizes the familiar spacetime translation, and it can be viewed as the angle-dependent translation.
It is related to the memory effect.
That is, the gravitational vacuum is actually degenerate, and the transition between vacuum states is parameterized by the supertranslation \cite{Strominger2014bms,Strominger:2014pwa,Strominger:2018inf}. 
Moreover, the memory effect can be viewed as the change in the radiative modes between two vacuum states. 
The vacuum transition is caused by the energy flux penetrating the null infinity, which is  conjugate to the supertranslation in the Hamiltonian formalism \cite{Flanagan:2015pxa}.
In fact, the amount of the null memory effect is proportional to the total energy radiated \cite{Strominger2014bms,Flanagan:2015pxa}.
The memory effect, supertranslation, and (leading) soft graviton theorem \cite{Weinberg:1965s} are three corners of the so-called infrared triangle, a triangular equivalence relation in the infrared regime \cite{Strominger:2018inf}.

Recently, the spin and center-of-mass (CM) memory effects were discovered \cite{Pasterski:2015tva,Nichols:2018qac}.
The effect discussed previously is thus specifically called the (leading) displacement memory effect.
The spin memory effect refers to the accumulated time delay between two photons propagating in the same orbit but in the opposite directions \cite{Himwich:2019qmj}.
Both of these novel effects could also be detected as the subleading displacement memory effect by interferometers \cite{Tahura:2020vsa,Hou:2021oxe}.
At the same time, the BMS group could be enlarged.
One may replace the Lorentz group by the Virasoro group to obtain the extended BMS group \cite{Barnich:2009se,Barnich:2010eb,Barnich:2011ct}.
Or, the Lorentz group could be generalized to the diffeomorphism group of the 2-sphere, and the asymptotic symmetry group is named generalized BMS group \cite{Campiglia:2014yka,Campiglia:2015yka,Campiglia:2020qvc}.
Let us call the new elements in the enlarged algebra the super-Lorentz generators. 
So the super-Lorentz generator refers to the vector in the Virasoro algebra in the case of the extended BMS group, or the vector generating the generic diffeomorphism on a 2-sphere in the case of the generalized BMS group.
We would like also to define super-rotation generator as the magnetic-parity component of the super-Lorentz generator, while the electric-parity component is named super-boost \cite{Compere:2018ylh}.
In either case, the ordinary spatial rotation is extended to the so-called super-rotation, which is related to the spin memory effect \cite{Kapec:2014opa,Campiglia:2014yka,Pasterski:2015tva,Nichols:2017rqr,Pasterski:2019ceq}, and the ordinary Lorentz boost is replaced by the super-boost, which is associated with the CM memory effect \cite{Nichols:2018qac}.
The triangular equivalence between the spin memory effect, the super-rotation and the subleading soft graviton theorem has also been established \cite{Kapec:2014opa,Cachazo:2014fwa,Campiglia:2014yka,Pasterski:2015tva,Strominger:2018inf,Pasterski:2019ceq}.
In the following, we will denote the enlarged BMS symmetry by $\overline{\text{BMS}}$, and as disclosed below, we follow the method of Ref.~\cite{Blanchet:2020ngx} to perform the asymptotic analysis, so we use $\ebms$ to refer to the generalized BMS group.

In addition to the memory effects mentioned above, there could exist infinite towers of memories, that are associated with infinite towers of residual gauge transformations in the harmonic gauge \cite{Compere:2019odm,Mao:2020vgh}.
Apart from the leading and some overleading gauge transformations, these residual gauge generators start at higher powers in a suitably defined radial coordinate, becoming trivial at the null infinity.
The equivalences among the memories, gauge symmetries, and infinite towers of soft theorems were also uncovered \cite{Compere:2019odm}.
Asymptotic analyses have been done for  higher dimensional spacetimes \cite{Hollands:2003ie,Hollands:2004ac,Tanabe:2011es,Kapec:2015vwa,Hollands:2016oma,Pate:2017fgt,Mao:2017wvx,Satishchandran:2019pyc}.
Studies showed that with appropriate boundary conditions on the metric and Ricci tensor components, there also exists nontrivial asymptotic symmetry group other than the asymptotic Poincar\'e group \cite{Awada:1985by,Kapec:2015vwa,Pate:2017fgt,Campiglia:2017xkp,Mao:2017wvx}.
However, these symmetry operations act on overleading terms in the metric expansion, while memory effects are described by relatively subleading terms. 
In this work, we consider the case of 4 dimension, and focus on the displacement, spin and CM memories.

In the modified theories of gravity, there might also exist the memory effect, as any alternatives to GR shall predict the existence of the tensor GWs by the most recent observations \cite{Abbott:2016blz,TheLIGOScientific:2017qsa,LIGOScientific:2021qlt,NANOGrav:2023gor,Reardon:2023gzh,Xu:2023wog}.
Indeed, there have been several works on the memories in some modified theories of gravity, such as Brans-Dicke (BD) theory \cite{Brans:1961sx,Hou:2020tnd,Tahura:2020vsa,Hou:2020wbo,Hou:2020xme,Tahura:2021hbk,Seraj:2021qja} and the dynamical Chern-Smions (dCS) theory \cite{Jackiw:2003pm,Alexander:2009tp,Hou:2021oxe,Hou:2021bxz}.
These works revealed similar memory effects discussed so far. 
The asymptotic symmetry of these theories is also the $\ebms$ symmetry, and the memories are related to the symmetry similarly as in GR \cite{Hou:2020tnd,Tahura:2020vsa,Hou:2021oxe}.
There are new memories associated with the extra degrees of freedom provided by the modified theories, called the scalar memory effects, as they are excited by the extra scalar degrees of freedom in these theories.
It is also interesting to note that although there are displacement, spin, and CM effects in these modified gravities, their magnitudes are actually different from those in GR, as the extra scalar degrees of freedom contribute to these memories.
This allows to use the memory effect to probe the nature of gravity.
There are more works on memory effects in modified gravities, such as \cite{Lang:2013fna,Lang:2014osa,Du:2016hww,Koyama:2020vfc,Kilicarslan:2018bia,Kilicarslan:2018yxd,Kilicarslan:2018unm}. 
In particular, Ref.~\cite{Heisenberg:2023prj} studied the tensor null memory effect in the dynamical metric theories using the ``high-frequency approximation'' developed by Isaacson \cite{Isaacson:1967zz,Isaacson:1968zza}.
It was found out, quite generally, that the tensor null memory effect is sourced by the null radiation of all the degrees of freedom in the theory.

One should note that both BD and dCS respect the local Lorentz invariance. 
Although dCS is said to violate the parity \cite{Zhao:2019xmm}, it occurs at the higher orders in the inverse of the radial coordinate, so it does not affect the memory effect \cite{Hou:2021oxe}.
In this work, we would like to consider yet another modified theory of gravity, Einstein-\ae{}ther theory \cite{Jacobson:2000xp}.
It is known as a local Lorentz-violating theory, as it possesses a nowhere vanishing, timelike vector field $\mfu^\mu$, called the \ae{}ther field.
This field thus defines a preferred reference of frame at each spacetime event, in which $\mfu^\mu$ is at rest. 
The local Lorentz invariance is thus spontaneously broken, once one chooses a suitable ``vacuum'' configuration for the \ae ther field.
Both the metric field $g_{\mu\nu}$ and the \ae ther field $\mfu^\mu$ mediate the gravitational interaction in this theory.
The GW solutions have been sought for in the flat spacetime background \cite{Jacobson:2004ts,Yagi:2013ava,Gong:2018cgj,Hou:2018djz,Lin:2018ken,Zhang:2019iim}.
These linearized analyses showed that there are 5 radiative degrees of freedom, including the tensor, vector, and scalar modes. 
These modes are allowed to propagate at the superluminal speeds, due to the breaking of the local Lorentz invariance.
Therefore, one expects this theory would predict new phenomena regarding memory effects.
That is, there are new vector and scalar memories in addition to the familiar tensor memories.
These memories might be related to the asymptotic symmetry in a novel way.

The main task of this work is to perform the asymptotic analysis for Einstein-\ae ther theory, and determine the asymptotic symmetries of the asymptotically flat spacetime. 
Memory effects will be identified, and associated with the asymptotic symmetries.
For these purposes, it would be better to find a suitable coordinate system, like Bondi-Sachs coordinates \cite{Bondi:1962px,Sachs:1962wk} or Newman-Unti coordinates \cite{Newman:1963ugj} used for analyzing asymptotically spacetime in the null direction in GR.
Suitable boundary conditions shall also be imposed on the dynamical fields, such that general solutions to the equations of motion can be solved for in these coordinates.
These boundary conditions shall not be too restrictive, otherwise, interesting solutions might be excluded.
The conditions may not be too weak, either, as too many solutions are permitted, and some of them might not seem to be flat in the large enough distances.
It is a delicate task to choose suitable boundary conditions, which has not yet been done systematically for Einstein-\ae ther theory.
Since this theory is greatly complicated, one may start with the linearized version.

In fact, even in the linearized theory, there are still some obstacles.
As mentioned above, there are 5 propagating degrees of freedom, satisfying d'Alembertian equations with different speeds. 
This would require us to analyze the behavior of each mode separately, as they will eventually arrive at different spacetime regions in the infinite future.
Moreover, due to the absence of the gravitational Cherenkov radiation \cite{Elliott:2005va}, these speeds shall be no less than 1.
So it seems that one shall employ drastically different methods used in GR to perform the asymptotic analysis for Einstein-\ae ther theory.
For example, naively, one would expect these GWs eventually arrive at the spatial infinity, so one would like to analyze the asymptotic behaviors of these waves using either the Ashtekar-Hansen formalism \cite{Ashtekar:1978zz} or Beig-Schmidt formalism \cite{Beig:1983grg,Beig:1984int}. 
However, these formalisms were presumably designed to investigate the nonradiative modes at the spatial infinity, while, here, the radiative modes of Einstein-\ae ther theory are to be studied.
So neither of these formalisms might be suitable.
Luckily, there actually exists an interesting field redefinition $(g_{\mu\nu},\mfu^\mu)\rightarrow(g'_{\mu\nu},\mfu'^\mu)$ \cite{Eling:2006ec}, such that there is at least one mode traveling at the speed 1, measured by a properly redefined metric field $g'_{\mu\nu}$.
Therefore, one may adapt the methods presented in Refs.~\cite{Blanchet:1986dk,Blanchet:2020ngx} for GR to analyze the asymptotic behavior of this specific mode.
Perform a different field redefinition, then, another mode would propagate at the speed 1, and its asymptotic behavior can be studied similarly.
In this way, one can determine the asymptotic behaviors of all radiative modes, and the associated asymptotic symmetries.

In Ref.~\cite{Blanchet:2020ngx}, the authors sought for the coordinate transformation that transforms the metric in the harmonic gauge $(t,x^j)$ to the Newman-Unti gauge $(\tdu,\tdr,\tdth^a)$ \cite{Newman:1963ugj}.
The leading order part (in $1/\tdr$) of the transformation can be freely specified, and thus defines the infinitesimal $\ebms$ transformation.
In this work, similar method will be adapted to Einstein-\ae ther theory. 
As discussed above, we would study the asymptotic behaviors of the radiative modes separately. 
For each mode, the redefined metric perturbation will be written in a certain gauge, and then, transformed to a suitable coordinate system, named pseudo-Newman-Unti coordinates.
This is because, in general, the radiative modes cannot be expressed in the standard Newman-Unti gauge in Einstein-\ae ther theory. 
If one insisted on using the standard Newman-Unti coordinates, the redefined metric excited by the vector or scalar modes would acquire terms proportional to $\ln \tdr$ or even $\tdr\ln \tdr$, relative to the respective leading parts.
As we are working in the linearized regime, these blowing up terms would exceed the leading Minkowski metric at the large enough distance, but this is inconsistent with the linearization.
Although there are polyhomogeneous solutions in GR, the metric components contain terms proportional to $r^{-n}\ln^kr$ with $n,k>0$ \cite{Andersson:1993we,Chrusciel:1993hx}.
So in this work, one seeks for a coordinate system, in which the coordinate $\tdr$ is nearly a null direction, $\tdu$ nearly a retarded time, and at the same time, there are no logarithmically diverging terms in the metric components.
In these coordinates, the redefined metric and the \ae ther field have no logarithmically diverging components, and behave well at the large $\tdr$.
We call such a kind of coordinate system the pseudo-Newman-Unti coordinate system.
Once the suitable boundary conditions are imposed on the redefine metric components, the asymptotic symmetries can be identified. 
As analyzed in the main text, the asymptotic symmetry group in Einstein-\ae ther theory includes the familiar $\ebms$ group as its subgroup.
Moreover, the boundary conditions allow the existence of a new symmetry, generated by a vector field $Z^a(\tdth^b)$ tangent to the 2-sphere.
$Z^a$ is subleading relatively to the super-Lorentz generator, so the symmetry generated by it will be called the subleading $\ebms$ symmetry.
Therefore, the asymptotic symmetry of Einstein-\ae ther theory includes the $\ebms$ symmetry and the subleading $\ebms$ symmetry.

The memory effects excited by the radiative modes will also be determined by integrating the geodesic deviation equations at $\tdr\rightarrow\infty$.
The relation between the memory effect and the asymptotic symmetry will be discussed. 
It turns out that for the tensor, vector and scalar degrees of freedom, one can identify their respective displacement, and the subleading displacement memory effects, given by relevant terms in the integrated geodesic deviation equations.
It is also possible to split the subleading displacement memory effect into the spin and CM memory parts in the tensor sector.
The tensor displacement memory effect shares many characteristics with the one in GR \cite{Strominger:2014pwa,Strominger:2018inf}, BD \cite{Hou:2020tnd,Hou:2020xme} and dCS \cite{Hou:2021oxe}.
For example, it can be viewed as the vacuum transition in the tensor sector, parameterized by a supertranslation, and of the electric-parity type. 
The vector displacement memory is intimately related to the subleading $\ebms$ symmetry, and unlike the tensor displacement memory or the velocity kick memory effect in electromagnetism \cite{Strominger:2018inf}, it has both the electric-parity and magnetic-parity components.
The scalar memories may have no explicit relation with the spacetime asymptotic symmetries, which also happens in other modified theories of gravity \cite{Hou:2020tnd,Tahura:2020vsa,Seraj:2021qja,Hou:2021oxe}.
Since only the linearized theory is considered in this work, one cannot obtain the constraint equations for the various memory effects, that are useful for calculating the magnitudes of the memories.
These constraint equations shall be derived once the nonlinear analysis is performed in the future work.

This work is organized as follows.
Section~\ref{sec-nc} collects notation and conventions.
In Section~\ref{sec-pert-eqs}, the basics of Einstein-\ae ther theory is reviewed, and the linearized equations of motion are obtained using the gauge-invariant formalism \cite{Flanagan:2005yc}.
Section~\ref{sec-sch} discusses the general scheme to construct the pseudo-Newman-Unti coordinates, and to identify the memory effects for each radiative mode.
Then, one starts with the construction of the pseudo-Newman-Unti coordinates for the tensor modes in Section~\ref{sec-pnu-t}.
There, one first determines the multipolar solution to the linearized equation of motion for the tensor mode in Section~\ref{sec-ten-vac-sols}, and then, fixes a suitable gauge condition for the redefined metric perturbation in Section~\ref{sec-t-gf}.
With these, one can find the pseudo-Newman-Unti coordinates, and the asymptotic symmetry will be discussed in Section~\ref{sec-psu-t-c}.
The memory effects are discussed in Section~\ref{sec-ten-mm}.
This procedure will be repeated for the vector and scalar modes in sections~\ref{sec-pnu-v} and \ref{sec-pnu-s}, respectively.
Finally, we will discuss the results and conclude in Section~\ref{sec-con}.

\subsection{Notation and conventions}
\label{sec-nc}

There are several coordinate systems used in this work. 
The pseudo-global Lorentz coordinates are denoted as $x^\mu=(t,x^j)$, and the associated spherical ones as $(t,r,\theta^a)$ with $\theta^a=(\theta,\varphi)$, and $r=|\vec x|$.
As one can see, letters $j,k,...,z$ are the space indices in the Cartesian coordinate system, and will be raised or lowered using $\delta^{ij}$ or its inverse, respectively. 
$a,b,...,h$ are the indices on the 2-dimensional sphere, and one uses $\gamma_{ab}$, the metric on the unit 2-sphere, and its inverse to lower and raise these indices.
On the unit 2-sphere, one has the natural basis $e_a=\pd/\pd\theta^a$, whose components in the Cartesian coordinates are $e^j_a=\pd n^j/\pd \theta^a$ with $n^j=x^j/r$.
It is easy to check that $n_je_a^j=0$.
Also, one has $\gamma_{ab}=\delta_{jk}e^j_ae_b^k$, $\pd_j\theta^a=r^{-1}\gamma^{ab}e_b^j$, and $\gamma^{ab}e_a^je_b^k\equiv\perp^{jk}=\delta^{jk}-n^jn^k$.
Let $\mathscr D_a$ be the covariant derivative compatible with $\gamma_{ab}$ with the properties $\mathscr D_ae_b^j=\mathscr D_be_a^j=\mathscr D_a\mathscr D_bn^j=-\gamma_{ab}n^j$ \cite{Blanchet:2020ngx}.
The pseudo-Newman-Unti coordinates are $\tilde x^\mu=(\tilde u,\tdr,\tdth^a)$, and the related ``Lorentz coordinates'' are $(\tilde t,\tilde x^j)$, which are associated with $\tilde x^\mu$ in the usual manner when the spacetime is flat.
The components of any tensor expressed in the pseudo-Newman-Unti coordinates are tilded, while those in the pseudo-global Lorentz coordinates are not.
The multi-index notation will be used. 
So $L$, as a subscript, means $j_1j_2\cdots j_l$.
In particular, $\pd_L=\pd_{j_1}\pd_{j_2}\cdots\pd_{j_l}$, and $n_L=n_{j_1}n_{j_2}\cdots n_{j_l}$.
Note that these are written in the pseudo-global Lorentz coordinates, and in the pseudo-Newman-Unti coordinates they are $\tpd_L=\tpd_{j_1}\tpd_{j_2}\cdots\tpd_{j_l}$, and $\tdn_L=\tdn_{j_1}\tdn_{j_2}\cdots\tdn_{j_l}$.
We will use the units such that $c=1$.

\section{Einstein-\AE ther Theory}
\label{sec-pert-eqs}

The action of Einstein-\ae ther theory is given by \cite{Jacobson:2004ts}
\begin{equation}\label{aeact}
\begin{split}
  S_{\text{EH-\ae}} =&\frac{1}{16\pi G} \int\ud^4x\sqrt{-g}[R-c_1(\nabla_\mu \tae_\nu)\nabla^\mu \tae^\nu\\
  &-c_2(\nabla_\mu \tae^\mu)^2-c_3(\nabla_\mu \tae_\nu)\nabla^\nu \tae^\mu\\
  &+c_4(\tae^\rho\nabla_\rho \tae^\mu)\tae^\sigma\nabla_\sigma \tae_\mu+\lambda(\tae^\mu \tae_\mu+1)],
  \end{split}
\end{equation}
where $\lambda$ is a Lagrange multiplier and $G$ is the gravitational constant,
the coupling constants $c_i\,(i=1,2,3,4)$ are expected to be of the order unity.
This theory is diffeomorphism invariant.
In general, $\mfu^\mu$ possesses no internal symmetries, in contrast to the 4-potential in Maxwell's electrodynamics \cite{Jackson:1998nia}. 
But since $\mfu^\mu$ couples with $g_{\mu\nu}$ non-minimally, one may study the symmetry properties of the metric to define the symmetry of $\mfu^\mu$, as shown in the later sections.
The Lagrange multiplier $\lambda$ forces $\tae^\mu$ a normalized timelike vector field. 
So $\tae^\mu$ defines a preferred reference frame at each spacetime point, and 
the local Lorentz invariance is thus spontaneously broken.

Ignoring the matter sector of the action, one can calculate the equations of motion, given by \cite{Gong:2018cgj},
\begin{subequations}
  \label{eq-eoms}
\begin{gather}
  R_{\mu\nu}-\frac{1}{2}g_{\mu\nu}R= T_{\mu\nu}^{\ae},\label{eq-ae-1}\\
  c_1\nabla_\mu\nabla^\mu \tae_\nu+c_2\nabla_\nu\nabla_\mu \tae^\mu+c_3\nabla_\mu\nabla_\nu \tae^\mu\nonumber\\
   -c_4  \nabla_\mu(\tae^\mu \mfa_\nu)+c_4\mfa_\mu\nabla_\nu \tae^\mu+\lambda \tae_\nu= 0,\label{eq-ae-2}\\
   \tae^\mu \tae_\mu+1=0,\label{eq-ae-3}
\end{gather}
where $\mfa^\mu=\tae^\nu\nabla_\nu \tae^\mu$ is the 4-acceleration of $\tae^\mu$, and the \ae ther stress-energy tensor $T_{\mu\nu}^{\ae}$ is
\begin{widetext}
\begin{eqnarray}
  T_{\mu\nu}^{\ae} &=& \lambda[\tae_\mu \tae_\nu-\frac{1}{2}g_{\mu\nu}(\tae^\rho \tae_\rho+1)]+c_1[(\nabla_\mu \tae_\rho)\nabla_\nu \tae^\rho-(\nabla_\rho \tae_\mu)\nabla^\rho \tae_\nu+\nabla_\rho(\tae_{(\mu}\nabla^\rho \tae_{\nu)}
  \nonumber\\
  &&-\tae_{(\mu}\nabla_{\nu)}\tae^\rho+\tae^\rho\nabla_{(\mu}\tae_{\nu)})]+c_2g_{\mu\nu}\nabla_\rho(\tae^\rho\nabla_\sigma \tae^\sigma)+c_3\nabla_\rho(\tae_{(\mu}\nabla_{\nu)}\tae^\rho-\tae_{(\mu}\nabla^\rho \tae_{\nu)}
  \nonumber\\
  &&+\tae^\rho\nabla_{(\mu}\tae_{\nu)})+c_4[\mfa_\mu \mfa_\nu-\nabla_\rho(2\tae^\rho \tae_{(\mu}\mfa_{\nu)}-\mfa^\rho \tae_\mu \tae_\nu)]
  \nonumber\\
  &&+\frac{1}{2}g_{\mu\nu}[-c_1(\nabla_\rho \tae_\sigma)\nabla^\rho \tae^\sigma-c_2(\nabla_\rho \tae^\rho)^2-c_3(\nabla_\rho \tae_\sigma)\nabla^\sigma \tae^\rho+c_4\mfa_\rho \mfa^\rho].
\end{eqnarray}
\end{widetext}
\end{subequations}
Here, Eq.~\eqref{eq-ae-3} is a constraint equation.

In the following, the GW solution will be sought for around the flat spacetime background, following Ref.~\cite{Gong:2018cgj}. 
The zeroth order solution is given by
\begin{equation}\label{0thsol}
  g_{\mu\nu}=\eta_{\mu\nu},\quad \mfu^\mu=\underline{\mfu}^\mu=(1,0,0,0),\quad\lambda=\underline\lambda=0.
\end{equation}
By analogue to the treatment of the spontaneous symmetry breaking in quantum field theory \cite{Schwartz:2013pla}, one may now perform an infinitesimal coordinate transformation $x^\mu\rightarrow x^\mu+\zeta^\mu$, which transforms the background metric and \ae ther fields in the following way,
\begin{gather}
\eta_{\mu\nu}\rightarrow\eta_{\mu\nu}-\partial_\mu\zeta_\nu-\partial_\nu\zeta_\mu,\\
\underline{\mfu}^\mu\rightarrow\underline{\mfu}^\mu+\partial_t\zeta^\mu,\label{eq-vev-tf}
\end{gather}
where, as usual, one uses $\eta_{\mu\nu}$ and $\eta^{\mu\nu}$ to lower and raise the greek indices from now on.
Therefore, as long as $\zeta^\mu$ depends on $t$, the vacuum expectation value (vev) $\underline{\mfu}^\mu$ is changed.
Such kind of $\zeta^\mu$ includes the Lorentz boosts ($\zeta^\mu=\varpi^\mu{}_tt$ with $\varpi^\mu{}_t$ constant and $\varpi^t{}_t=0$) as the special cases, and thus, the Lorentz symmetry is spontaneously broken by $\underline{\mfu}^\mu$.
Of course, $\zeta^\mu$ can be more general than the Lorentz symmetry generators, so the ``generalized Lorentz boost symmetry'', or ``superboost symmetry'' generated by a time-dependent vector field $\zeta^\mu$ is broken by $\underline{\mfu}^\mu$.

Now, perturb the metric and the \ae ther field in the following way,
\begin{subequations}
\label{eq-pbs}
\begin{gather}
  g_{\mu\nu}=\eta_{\mu\nu}+h_{\mu\nu},\\
  \mfu^\mu=\underline{\mfu}^\mu+\mfv^\mu.\label{eq-u-pb}
\end{gather}
\end{subequations}
And of course, $\lambda$ shall be  treated to be of the same order as $h_{\mu\nu}$ and $\mfv^\mu$.
Comparing Eqs.~\eqref{eq-vev-tf} and \eqref{eq-u-pb}, one basically promotes the gauge transformation parameters $\zeta^\mu$ to dynamical fields $\mfv^\mu$, similarly to the treatment of the Higgs mechanism \cite{Srednicki:2007qs,Schwartz:2013pla}.  
In fact, the \ae{}ther field plays a role of the Higgs field in this theory.
Substituting Eq.~\eqref{eq-pbs} into the equations of motion \eqref{eq-eoms}, and keeping the linear terms in the field perturbations, one obtains the linearized equations of motion, which are too complicated to be explicitly written down. 
To simplify the linearized equations, one makes use of the gauge-invariant formalism \cite{Flanagan:2005yc,Gong:2018cgj}.
To this end, one first decomposes the metric perturbation $h_{\mu\nu}$ and the perturbed \ae ther field $\mfv^\mu$ as,
\begin{subequations}
 \begin{gather}
  h_{tt} = 2\phi, \label{httdec}\\
  h_{tj} = \beta_j+\partial_j\gamma, \label{htjdec}\\
  h_{jk} = h_{jk}^\mathrm{TT}+\frac{H}{3}\delta_{jk}+\partial_{(j}\varepsilon_{k)}+\left(\partial_j\partial_k-\frac{\delta_{jk}}{3}\nabla^2\right)\rho,\label{hjkdec}\\
  \mfv^t=\frac{1}{2}h_{tt}=\phi,\label{v0dec}\\
   \mfv^j=\mu^j+\partial^j\omega.\label{vjdec}
\end{gather}
\end{subequations}
In the above expressions, $h_{jk}^\mathrm{TT}$ is the transverse-traceless part of $h_{jk}$, satisfying $\pd^kh_{jk}^\mathrm{TT}=0$ and $\delta^{jk}h_{jk}^\mathrm{TT}=0$. 
$\beta_j,\,\varepsilon_j$ and $\mu^j$ are transverse vectors.
Equation~\eqref{v0dec} is due to $\mfu^\mu\mfu_\mu=-1$.
Several gauge-invariant variables can be defined, which are
\begin{subequations}
\begin{gather}
h_{jk}^\mathrm{TT},\\
  \Phi = \dot\gamma-\phi-\frac{\ddot\rho}{2}, \;\;\;
  \Psi = \frac{1}{3}(H-\nabla^2\rho), \;\;\;
   \Omega=\omega+\frac{\dot\rho}{2},\\
  \Xi_j = \beta_j-\frac{1}{2}\dot\varepsilon_j,\quad
  \Sigma_j=\beta_j+\mu_j.
\end{gather}
\end{subequations}
Using these gauge-invariant variables, one can rewrite the linearized equations of motion.
It turns out that some gauge-invariant variables satisfy the following d'Alembertian equations,
\begin{subequations}
  \label{eq-weqs}
\begin{gather}
  -\frac{1}{s_{\mathsf g}^2}\ddot h_{jk}^\mathrm{TT}+\nabla^2h_{jk}^\mathrm{TT}=0,\label{ettjk}\\
  -\frac{1}{s_{\mathsf v}^2}\ddot\Sigma_j+\nabla^2\Sigma_j=0,\label{eomsig} \\
  -\frac{1}{s_{\mathsf s}^2}\ddot\Omega+\nabla^2\Omega=0.\label{eomscl2}
\end{gather}
\end{subequations}
The squared speeds of these modes are
\begin{subequations}
\begin{gather}
  s_{\mathsf g}^2=\frac{1}{1-c_{+}},\label{tenspd}\\
  s_{\mathsf v}^2=\frac{2c_1-c_+c_-}{2c_{14}(1-c_{+})},\label{vecspd}\\
  s_{\mathsf s}^2=\frac{c_{123}(2-c_{14})}{c_{14}(1-c_{+})(2+2c_2+c_{123})},\label{sclspd}
\end{gather}
\end{subequations}
where $c_{\pm}=c_1\pm c_3$, $c_{14}=c_1+c_4$, and $c_{123}=c_1+c_2+c_3$.
The remaining gauge-invariant variables are given by
\begin{subequations}
\begin{gather}
  \Phi=\frac{c_{14}-2c_+}{2-c_{14}}\dot\Omega,\label{eq-s-1}\\
  \Psi=\frac{2c_{14}(c_+-1)}{2-c_{14}}\dot\Omega,\label{eq-s-2}\\
  \Xi_j=\frac{c_+}{c_+-1}\Sigma_j.\label{eq-xieom}
\end{gather}
\end{subequations}
These are dependent variables.
For the details of deriving the above results, please refer to Ref.~\cite{Gong:2018cgj}.
So there are only five propagating physical degrees of freedom.
Two of them are tensor modes (${\mathsf g}$), encoded in $h_{jk}^\mathrm{TT}$, another two are vector modes (${\mathsf v}$), given by $\Sigma_j$, and the remaining one is a scalar (${\mathsf s}$) degree of freedom  represented by $\Psi$.
These modes, collectively denoted by $\sfm(=\sfg,\sfv,\sfs)$, generally propagate at different speeds $s_\sfm$, which can be greater than, equal to, or less than the speed of light.
When $c_+=c_4=0$ and $2c_1c_2=c_2-c_1$, these speeds are all one.

These radiative modes can be related to the GW polarizations using the geodesic deviation equation \cite{Wald:1984rg},
\begin{equation}
  \label{eq-gdv}
  \frac{\ud^2S^j}{\ud \tau^2}\approx-R_{tjtk}S^k,
\end{equation}
for two test particles with the affine parameter $\tau$, separated by a small deviation vector $S^j$.
In this equation, $R_{tjtk}$ is called the electric part of the Riemann tensor, expressed in terms of the gauge-invariant variables as \cite{Flanagan:2005yc,Gong:2018cgj}
\begin{equation}
  \label{eq-rtjtk}
  R_{tjtk}=-\frac{1}{2}\ddot h_{jk}^\text{TT}+\pd_{(j}\dot\Xi_{k)}+\pd_{jk}\Phi-\frac{1}{2}\ddot\Psi\delta_{jk}.
\end{equation}
To identify the GW polarizations, one simple method is to solve for the plane wave solutions to Eq.~\eqref{eq-weqs}, assuming the GWs propagate in the positive $z$ direction.
Then, after substituting these solutions to Eq.~\eqref{eq-rtjtk}, one finds out that $-R_{txtx}+R_{tyty}=\ddot h_{xx}^\text{TT}$ and $R_{txty}=-\ddot h_{xy}^\text{TT}$.
So the tensor modes excite the plus and cross polarizations.
In addition, $R_{txtz}\propto\partial_z\dot\Sigma_x$ and $R_{tytz}\propto\partial_z\dot\Sigma_y$ with the same proportionality factor, which means that the vector modes excite the vector-$x$ and vector-$y$ polarizations.
And finally, $R_{txtx}+R_{tyty}\propto \dddot\Omega$ and $R_{tztz}\propto\dddot\Omega$ with different proportionality factors, and thus, the scalar mode excites the longitudinal and the breathing polarizations.
For more gauge-invariant ways of identifying the polarizations, please refer to Ref.~\cite{Alves:2023rxs}.

As shown above, the propagating speeds of the radiative modes are generally different.
They will propagate to different spacetime regions eventually.
This requires us to study their asymptotic behaviors separately. 
In addition, their speeds are not necessarily equal to one, which makes the asymptotic analysis even more complicated. 
Thanks to Ref.~\cite{Eling:2006ec}, one can rewrite the action~\eqref{aeact} with newly defined quantities given by,
\begin{equation}
  \label{eq-def-disf}
  g'_{\mu\nu}=g_{\mu\nu}+(1-\sigma)\mfu_\mu \tae_\nu,\quad\tae'^\mu=\frac{\tae^\mu}{\sqrt{\sigma}},
\end{equation}
where $\sigma$ is a real constant and can be chosen for convenience,
then the action would take the same form with the coupling constants $c_i$'s transforming in the way given by Eqs. (15) -- (18) in Ref.~\cite{Eling:2006ec}.
This observation will be very useful for the following discussion. 
It implies that if $g_{\mu\nu}$ possesses a Killing vector field $K^\mu$, so does $g'_{\mu\nu}$.
In fact, one can show that for any vector field $v^\mu$, 
\begin{equation}
  \label{eq-lie-gr}
  \lie_vg'_{\mu\nu}=2\nabla_{(\mu}v_{\nu)}+2(1-\sigma)\mfu_{(\mu}\lie_v\mfu_{\nu)}=2\nabla'_{(\mu}v'_{\nu)},
\end{equation}
where $\nabla'_{\mu}$ is the covariant derivative associated with $g'_{\mu\nu}$, and $v'_\mu=g'_{\mu\nu}v^\nu$.
Suppose $v^\mu=K^\mu$, and let $\psi_{\iota}$ be the diffeomorphism generated by $K^\mu=(\ud/\ud\iota)^\mu$ with $\iota$ a parameter for the integral curves of $K^\mu$.
Then, in order that $\psi^*_{\iota}g_{\mu\nu}(=g_{\mu\nu})$, $\psi_{\iota}^*\mfu^\mu$, and $\psi^*_{\iota}\lambda$ still satisfy the equations of motion \eqref{eq-eoms}, there should be $\psi_{\iota}^*\mfu^\mu=\mfu^\mu$ and $\psi^*_{\iota}\lambda=\lambda$.
Therefore, $\lie_K\mfu^\mu=0$, that is, $\lie_K\mfu_\mu=\lie_K(g_{\mu\nu}\mfu^\nu)=0$ \cite{Wald:1984rg}.
Then, by Eq.~\eqref{eq-lie-gr}, $\lie_K g'_{\mu\nu}=0$, which means that $g_{\mu\nu}$ and $g'_{\mu\nu}$ share the same symmetry.
For the current work, if $g_{\mu\nu}$ possesses the asymptotic symmetry, so does $g'_{\mu\nu}$.
One can work in the theory defined either by $(g_{\mu\nu},\tae^\mu)$, or by $(g'_{\mu\nu},\tae'^\mu,\sigma)$ to determine the asymptotic symmetry.
We say $(g_{\mu\nu},\tae^\mu)$ defines the ``physical'' frame, while $(g'_{\mu\nu},\tae'^\mu,\sigma)$ the ``unphysical'' frame.

An even more interesting implication of the redefinition~\eqref{eq-def-disf} is that the speeds of the radiative modes in the unphysical frame are given by \cite{Eling:2006ec}
\begin{equation}
  \label{eq-upvs}
  s'_\sfm=\frac{s_\sfm}{\sqrt\sigma}.
\end{equation}
So if one sets $\sigma=s_{\sfm'}^2$ , then the speed of $\sfm'$ is 1 in the unphysical frame.
For example, if we consider the asymptotic behaviors of the tensor mode ($\sfm'=\sfg$), one can set $\sigma=s^2_\sfg$, then in the unphysical frame, the tensor mode propagates at the speed $s'_\sfg=1$.
This may allow us to borrow the idea of Ref.~\cite{Blanchet:2020ngx} for GR to analyze the asymptotic behaviors of the radiative modes in Einstein-\ae ther theory.
The basic strategy is to first solve the linearized equations of motion for multipolar solutions in a convenient coordinate system, which can be easily done, and then find the coordinate transformation such that in a new coordinate system $(\tdu,\tdr,\tdth^a)$, the components of the metric and \ae ther fields are expressed as series expansions in $1/\tdr$ with the expansion coefficients depending on $(\tdu,\tdth^a)$.
As long as this coordinate system is determined, one can analyze the asymptotic behaviors of the metric and \ae ther fields, asymptotic symmetries, and memory effects.
This will be done explicitly in the following sections.
Since Einstein-\ae ther theory is quite different from GR, one may not reproduce the results in Ref.~\cite{Blanchet:2020ngx} as shown below.

Before preceding further, let us review some experimental constraints on Einstein-\ae ther theory very briefly.
For more complete discussions, please refer to more recent works~\cite{Gong:2018cgj,Oost:2018tcv,Gupta:2021vdj,Tsujikawa:2021typ}.
Ever since its birth, this theory has been constrained by several experimental observations.
There are bounds on the post-Newtonian parameters $\alpha_1$ and $\alpha_2$ parameterizing the local Lorentz violation \cite{Will:2014kxa}, the requirement that the GW carry positive energy \cite{Jacobson:2008aj}, and there should be no gravitational Cherenkov radiation \cite{Elliott:2005va}, etc.
Most recently, the observations of GW170817 and GRB 170817A set a strong constraint on the speed of the tensor mode, $-3\times10^{-15}\le s_\sfg-1\le7\times10^{-16}$ \cite{TheLIGOScientific:2017qsa,Goldstein:2017mmi,Savchenko:2017ffs,Monitor:2017mdv}.
Combining all these observations together, one can determine the constraints on Einstein-\ae ther theory, as discussed in Refs.~\cite{Gong:2018cgj,Oost:2018tcv,Gupta:2021vdj,Tsujikawa:2021typ,Ghosh:2021cub,Schumacher:2023cxh}.
However, in the following discussion, we would formally keep the coupling constants $c_i$'s free so that the results thus obtained are general enough. 
As soon as one fixes the values of $c_i$'s based on the experimental observations, one can substitute these values into expressions presented below to obtain the asymptotic behaviors of the metric and \ae ther fields.

\section{General scheme to construct pseudo-Newman-Unti coordinates}
\label{sec-sch}

As discussed above, different modes generally travel at different speeds, so they will arrive at different spacetime regions in the infinite future.
This means that one may want to analyze the asymptotic behaviors of the radiative modes, separately.
That is, one need obtain the desired coordinate system for each radiative mode $\sfm$, individually. 
Since the procedures for determining the coordinate systems are similar among the radiative modes, in this section, let us discuss the general scheme to construct the pseudo-Newman-Unti coordinates.

So let us now consider a specific mode $\sfm\,(=\sfg,\sfv$ or $\sfs)$.
The first step is to solve the linearized equation of motion \eqref{eq-weqs} for the mode $\sfm$.
As in Ref.~\cite{Blanchet:2020ngx}, one would like to look for the most general multipolar solutions, which can be written as series expansions in $1/r$.
Since Eqs.~\eqref{eq-weqs} are written in terms of gauge-invariant variables, one should now fix the gauge. 
For example, one can set $\varepsilon_j=0$, and $\omega=\gamma=0$ \cite{Foster:2006az}.
Then, one should transform to the unphysical frame with $\sigma=s_\sfm^2$ so that the particular mode $\sfm$ being considered has a unit speed. 
Let the unphysical metric and \ae ther fields be written as $g'^{(\sfm)}_{\mu\nu}$ and $\mfu'^\mu_{(\sfm)}$, respectively. 
One can check that they have the following components,
\begin{equation}
  \label{eq-g1-tt}
g'^{(\sfm)}_{tt}=-s^2_\sfm+\cdots,\quad \mfu'^{t}_{(\sfm)}=\frac{1}{s_\sfm }+\cdots,
\end{equation}
at the leading orders in $1/r$, with $\cdots$ representing higher order corrections.
It would be better to perform the following coordinate transformation,
\begin{equation}
  \label{eq-1st-ctf}
  t\rightarrow \bar t=s_\sfm t,\quad x^j\rightarrow x^j,
\end{equation}
such that in the new coordinates $(\bar t,x^j)$, $g'^{(\sfm)}_{\bar t\bar t}=-1+\cdots$, and $\mfu'^{\bar t}_{(\sfm)}=1+\cdots$.
This means that, in this new coordinates, the leading term of $g'^{(\sfm)}_{\mu\nu}$ takes the usual form, namely, $g'^{(\sfm)}_{\mu\nu}=\eta'_{\mu\nu}+\cdots$ with $\eta'_{\mu\nu}=\text{diag}(-1,1,1,1)$.
In fact, all symbols in this section and the following are for the mode $\sfm$, and it would be clearer if one appended the subscript $\sfm$ to them.
However, this would be very cumbersome.
So we will not append any subscript or superscript to any of these symbols, except $g'^{(\sfm)}_{\mu\nu}$, $\mfu'^\mu_{(\sfm)}$, and $\chi^\mu_{(\sfm)}$ defined in Eq.~\eqref{eq-def-chi}.
It should be easy to understand which mode these symbols are associated with based on the context.

Generally speaking, in this coordinates $(\bar t,x^j)$, the metric perturbation $h'_{\mu\nu}\equiv g'^{(\sfm)}_{\mu\nu}-\eta'_{\mu\nu}$, or rather its trace-reversed version $\bar h'_{\mu\nu}=h'_{\mu\nu}-\eta'_{\mu\nu}\eta'^{\rho\sigma}h'_{\rho\sigma}/2$, takes a form that would make the construction of the pseudo-Newman-Unti coordinates very complicated. 
So it is better to perform a further gauge transformation, parameterized by 
\begin{subequations}
  \label{eq-def-inf-gtf}
\begin{gather}
  \xi_{\bar t}=\sum_{l=0}\pd_L\frac{\sW_L}{r},\\
  \xi_j=\sum_{l=0}\pd_{jL}\frac{\sV_L}{r}+\sum_{l=1} \pd_{L-1}\left( \frac{\sT_{jL-1}}{r}+\epsilon_{jpq}\pd_{p}\frac{\sR_{qL-1}}{r} \right).
\end{gather}
\end{subequations}
Here, the components of $\xi^\mu$ are expressed in terms of the transverse-tracefree tensors $\sW_L$, $\sV_L$, $\sT_L$, and $\sR_L$, which are functions of $\bar t-r$.
Therefore, $\bar h'_{\mu\nu}$  and $\mfv'^\mu$ transform according to \cite{Carroll:2004st,Gong:2018cgj},
\begin{subequations}
  \label{eq-sim}
\begin{gather}
  \label{eq-gtf-h}
  \bar h'_{\mu\nu}\rightarrow\bar h''_{\mu\nu}=\bar h'_{\mu\nu}-\pd_\mu\xi_\nu-\pd_\nu\xi_\mu+\eta'_{\mu\nu}\pd_\rho\xi^\rho,\\
  \mfv''^\mu=\mfv'^\mu+\pd_{\bar t}\xi^\mu.
\end{gather}
\end{subequations}
One can choose suitable $\sW_L$, $\sV_L$, $\sT_L$, and $\sR_L$ for the mode $\sfm$ such that $\bar h''_{\mu\nu}$ takes a somewhat simple form.
One may call such a choice the \textit{good} gauge.
In GR, the good gauge is the transverse gauge $\pd_\nu\bar h^{\mu\nu}=0$ \cite{Blanchet:2020ngx}.

Now, one is ready to construct the pseudo-Newman-Unti coordinate system for the mode $\sfm$, by determining the appropriate coordinate transformation that transforms the metric perturbation $\bar h''_{\mu\nu}$.
Up to the linear order in the perturbations, one writes the coordinate transformation for the radiative mode $\sfm$ in the following way,
\begin{equation}
  \label{eq-def-ctf-2-nu}
  \tdu=u+\sfU,\quad \tdr=r+\sfR,\quad \tdth^a=\theta^a+\Theta^a,
\end{equation}
where $u=\bar t-r$.
$\sfU$, $\sfR$, and $\Theta^a$ are all linear in the field perturbations.
Before presenting the scheme to construct the pseudo-Newman-Unti coordinates, let us take a detour to review what the Newman-Unti gauge is and how it may be achieved by the coordinate transformations.

In the unphysical frame, the Newman-Unti coordinate system $(\tdu,\tdr,\tdth^a)$ is defined to be the one in which the metric $\tilde g'^{(\sfm)}_{\mu\nu}$ has the following components \cite{Newman:1962cia},
\begin{subequations}
\begin{equation}
  \label{eq-def-nu-up}
  \tilde g'^{(\sfm)}_{rr}=\tilde g'^{(\sfm)}_{ra}=0,\quad\tilde g'^{(\sfm)}_{ur}=-1.
\end{equation}
Or equivalently,
\begin{equation}
  \label{eq-def-nu-up-inv}
  \tilde g'^{uu}_{(\sfm)}=\tilde g'^{ua}_{(\sfm)}=0,\quad\tilde g'^{ur}_{(\sfm)}=-1.
\end{equation}
\end{subequations}
Thus, $\tdr$ is the null coordinate.
By Eq.~\eqref{eq-def-nu-up-inv}, one finds the conditions,
  \begin{gather*}
    g'^{\mu\nu}_{(\sfm)}\frac{\pd\tdu}{\pd x^\mu}\frac{\pd\tdu}{\pd x^\nu}=0,\\
    g'^{\mu\nu}_{(\sfm)}\frac{\pd\tdu}{\pd x^\mu}\frac{\pd\tdth^a}{\pd x^\nu}=0,\\
    g'^{\mu\nu}_{(\sfm)}\frac{\pd\tdu}{\pd x^\mu}\frac{\pd\tdr}{\pd x^\nu}=-1.
  \end{gather*}
Substituting Eq.~\eqref{eq-def-ctf-2-nu} into the above expressions, one obtains the following results \cite{Blanchet:2020ngx},
\begin{subequations}
  \begin{gather}
    k^\mu\pd_\mu\mathsf U=-\frac{1}{2} k_\mu k_\nu\bar h''^{\mu\nu},\\
    k^\mu\pd_\mu \sfR=\frac{1}{2}n_{jk}\bar h''_{jk}-\frac{1}{2}\bar h''_{jj}-\pd_{\bar t}\sfU,\\
    k^\mu\pd_\mu\Theta^a=\frac{e^a_j}{ r}\left( \pd_j\sfU+ k_\mu\bar h''^{\mu j} \right),
  \end{gather}
\end{subequations}
where $k^\mu=(1,n^j)$ is the leading order piece of $\tdr^\mu\equiv(\pd/\pd\tdr)^\mu$.
The left-hand sides of above equations can be viewed as the total derivative with respect to $r$, i.e., $k^\mu\pd_\mu\equiv\ud/\ud r$, and the right-hand sides are series expansions in $1/r$.
Once these set of equations are integrated, one can determine the remaining metric components using,
  \begin{gather*}
    \tilde g'^{rr}_{(\sfm)}=g'^{\mu\nu}_{(\sfm)}\frac{\pd\tdr}{\pd x^\mu}\frac{\pd\tdr}{\pd x^\nu},\\
    \tilde g'^{ra}_{(\sfm)}=g'^{\mu\nu}_{(\sfm)}\frac{\pd\tdr}{\pd x^\mu}\frac{\pd\tdth^a}{\pd x^\nu},\\
    \tilde g'^{ab}_{(\sfm)}=g'^{\mu\nu}_{(\sfm)}\frac{\pd\tdth^a}{\pd x^\mu}\frac{\pd\tdth^b}{\pd x^\nu}.
  \end{gather*}
Thus, $\tilde g'^{(\sfm)}_{ab}=[\tilde g'^{ab}_{(\sfm)}]^{-1}$, $\tilde g'^{(\sfm)}_{ua}=\tilde g'^{rb}_{(\sfm)}\tilde g'^{(\sfm)}_{ab}$, and $\tilde g'^{(\sfm)}_{uu}=-\tilde g'^{rr}_{(\sfm)}+\tilde g'^{ra}_{(\sfm)}\tilde g'^{(\sfm)}_{ua}$. 
Actually, this scheme can be applied to the tensor mode in Einstein-\ae{}ther theory, as in the unphysical frame defined by $\sigma=s_{\mathsf g}^2$, the tensor mode propagates at the speed of unity, and the metric perturbation $\bar h''_{\mu\nu}$ takes exactly the same form as in GR.
However, if one directly applies this scheme to the vector and scalar modes, one obtains terms proportional to $\ln\tdr$ or even $\tdr\ln\tdr$, relative to the respective leading order terms.
These terms would make the leading order part of the metric field $g'^{(\sfm)}_{\mu\nu}$, i.e., the Minkowski metric $\eta'_{\mu\nu}$, shadowed by the higher term part $h'_{\mu\nu}$,  as $\tdr\rightarrow+\infty$, which is inconsistent with the linearization. 
Since one has performed the transformation \eqref{eq-gtf-h} to take the good gauge to simplify the calculation, one may argue that it is probable to impose the Newman-Unti gauge condition without introducing any diverging logarithmic terms as one can execute a different gauge transformation. 
As a matter of fact, this is impossible, as one can do the same calculation after performing a further gauge transformation.
The diverging logarithmic terms always survive the gauge transformation \eqref{eq-gtf-h}.
Therefore, the Newman-Unti coordinate system may not be suitable for studying the vector and scalar modes.

The appearance of $\ln\tdr$ and $\tdr\ln\tdr$ terms comes from the rather strong requirements~\eqref{eq-def-nu-up} or \eqref{eq-def-nu-up-inv}.
These equations imply that $\tdr$ is a null direction from some finite place in the bulk all the way to the infinity $\tdr\rightarrow +\infty$.
Since in this work, one is interested in the fields at the infinity, one may relax these requirements, and demand $\tilde g'^{(\sfm)}_{rr}\sim\tilde g'^{(\sfm)}_{ur}+1\sim\order{1/\tdr}$ and $\tilde g'^{(\sfm)}_{ra}\sim\order{\tdr^0}$.
In this way, although $\tdr$ is not null everywhere, it is null at the infinity.
So based on this argument, let us determine the desired coordinate transformation.
After some tedious calculation, one knows that in the pseudo-Newman-Unti coordinates, the unphysical metric is given by
\begin{subequations}
  \label{eq-gc-pnu}
  \begin{gather}
    \tdg'^{(\sfm)}_{uu}=-1+2(\sfU^{[1]}+\sfR^{[1]})+\frac{1}{2}(\bar h''_{\bar t\bar t}+\bar h''_{jj}),\\
    \tdg'^{(\sfm)}_{ur}=-1+k^\mu\pd_\mu(\sfU+\sfR)+\sfU^{[1]}\nonumber\\
    +\frac{1}{2}(\bar h''_{tt}+2n_j\bar h''_{\bar tj}+\bar h''_{jj}),\label{eq-tdg-ur}\\
    \tdg'^{(\sfm)}_{ua}=-r^2\Theta_a^{[1]}+\mathscr D_a(\sfU+\sfR)+re_a^j\bar h''_{\bar tj},\\
    \tdg'^{(\sfm)}_{rr}=2k^\mu\pd_\mu\sfU+\bar h''_{\bar t\bar t}+2n_j\bar h''_{\bar tj}+n_{jk}\bar h''_{jk},\label{eq-tdg-rr}\\
    \tdg'^{(\sfm)}_{ra}=-r^2k^\mu\pd_{\mu}\Theta_a+\mathscr D_a\sfU+re_a^j(\bar h''_{\bar tj}+\bar h''_{jk}n_k),\label{eq-tdg-ra}\\
    \tdg'^{(\sfm)}_{ab}=\tdr^2\tdgamma_{ab}
    -\tdr^2 \left( 2\mathscr D_{(a}\Theta_{b)}+\frac{2}{\tdr}\sfR\tdgamma_{ab}\right.\nonumber\\
    \left.+\frac{1}{2}\tdgamma_{ab}\bar h''-e_a^je_b^k\bar h''_{jk} \right).\label{eq-tdg-ab-f}
  \end{gather}
\end{subequations}
As Eq.~(3.4c) in Ref.~\cite{Blanchet:2020ngx}, Eq.~\eqref{eq-tdg-ab-f} is explicitly written in terms of $\tdr$.
In the other equations, $r$ is used, instead of $\tdr$.
This is all right, as they are different from each other by $\sfR$, which is of the first order in the field perturbations.
Obviously, all terms in the above equations involving $\bar h''_{\mu\nu}$ are series expansions in $1/r$, without any logarithmic terms.
So in order for $\tdg'^{(\sfm)}_{\mu\nu}$ approaching the Minkowski metric at $r\rightarrow+\infty$, one has to impose certain conditions on $\sfU,\sfR$ and $\Theta^a$.

Since $\bar h''_{\mu\nu}$ are series expansions in $1/r$, it is natural to assume that
\begin{equation}
  \label{eq-sfU-se}
  \sfU=\sum_{l=0}\frac{\sfU_{\{l\}}}{r^l}+\sum_{l=1}r^l\sfU_{\{-l\}},
\end{equation}
where the positive powers of $r$ are included for generality.
$\sfR$ and $\Theta^a$ are also expanded in the similar manner.
This form of expansion is inspired by the usual Lorentz boost generator  whose radial component actually contains a positive power term \cite{Sachs:1962wk,Sachs:1962zza,Blanchet:2020ngx}.
It is also inspired by the form of the gauge transformation generator in the dual formalism of the scalar fields in some modified theories of gravity \cite{Seraj:2021qja,Hou:2021bxz}.
Whether the coefficients of the positive power terms exist or not depends on the boundary conditions imposed on the metric components.
These boundary conditions can be chosen as 
\begin{subequations}
  \label{eq-bdyc}
  \begin{gather}
    \tdg'^{(\sfm)}_{rr}\sim\order{\frac{1}{r}},\label{eq-bdyc-1-1}\\ \tdg'^{(\sfm)}_{ur}+1\sim\order{\frac{1}{r}},\label{eq-bdyc-1-2}\\ \tdg'^{(\sfm)}_{ra}\sim\order{r^0},\label{eq-bdyc-1-3}\\
\tdg'^{(\sfm)}_{uu}\sim\order{r^0},\quad\tdg'^{(\sfm)}_{ua}\sim\order{r^0},\label{eq-bdyc-2}\\
\det(\tdg'^{(\sfm)}_{ab})=\tdr^4\det(\tdgamma_{ab})+\order{\tdr^2}.\label{eq-bdyc-3}
  \end{gather}
\end{subequations}
Let us look into these conditions and their implications.
Firstly, Eq.~\eqref{eq-bdyc-1-1} guarantees $\tdr$ being null at the infinity, and results in $\sfU_{\{-l\}}=0$ for $l\ge1$. 
So there are no positive powers in $\sfU$, in fact.
Secondly, Eq.~\eqref{eq-bdyc-1-2} implies $\sfR_{\{-l\}}=0$ for $l\ge2$, so
\begin{equation}
  \label{eq-def-exp-r}
  \sfR=r\sfR_{\{-1\}}+\sfR_{\{0\}}+\sfR',
\end{equation}
with $\sfR'$ standing for the higher order terms in $1/r$,
and 
\begin{equation}
  \label{eq-dg-ur-ra}
  \pd_{\bar t}\sfU_{\{0\}}+\sfR_{\{-1\}}=0. 
\end{equation}
So the only surviving positive power term in $\sfR$ is $r\sfR_{\{-1\}}$, which  also exists in GR \cite{Blanchet:2020ngx}.
Thirdly, for $\tdg'^{(\sfm)}_{ra}$, there is no problem to require it to be $\order{r^0}$ as in Eq.~\eqref{eq-bdyc-1-3}, and this implies that $\Theta^a_{\{-l\}}=0$ for $l\ge1$.
Then, demand Eq.~\eqref{eq-bdyc-2}, leading to
\begin{gather}
  \pd_{\bar t}\sfR_{\{-1\}}=0,\label{eq-dg-uu}\\
  \pd_{\bar t}\Theta^a_{\{0\}}=0,\quad\pd_{\bar t}\Theta^a_{\{1\}}-\mathscr D^a\sfR_{\{-1\}}=0,\label{eq-dg-ua}
\end{gather}
respectively.
Finally, one wants Eq.~\eqref{eq-bdyc-3}, and so
\begin{equation}
  \mathscr D_a\Theta^a_{\{0\}}+2\sfR_{\{-1\}}=0.\label{eq-dg-det}
\end{equation}
This result is derived using $\det(\tdg'^{(\sfm)}_{ab})\sim\tdr^4\det(\tdgamma_{ab})$ at the leading order.
The condition that $\det(\tdg'^{(\sfm)}_{ab})$ vanishes at $\order{\tdr^3}$ leads to a relation involving $\sfR_{\{0\}}$ and field perturbations.
Since now we are presenting the general method to perform the asymptotic analysis, we do not give the explicit relation, which shall be displayed in the later sections. 
Up to now, one may solve Eqs.~\eqref{eq-dg-ur-ra}, \eqref{eq-dg-uu}, \eqref{eq-dg-ua}, and \eqref{eq-dg-det}, and get
\begin{subequations}
\begin{gather}
  \Theta^a_{\{0\}}=-Y^a(\theta^b), \\ 
  \sfR_{\{-1\}}=\frac{1}{2}\mathscr D_aY^a,\\
  \sfU_{\{0\}}=-f\equiv-T(\theta^a)-\frac{u}{2}\mathscr D_aY^a,\\
  \Theta^a_{\{1\}}=-Z^a(\theta^b)-\mathscr D^af,\label{eq-sfu0-th1-sol}
\end{gather}
where $T$, $Y^a$,  and $Z^a$ depend only on the angular coordinates.
\end{subequations}

Let us collect the results obtained so far, and define a vector field $\chi^\mu_{(\sfm)}$ with
\begin{subequations}
  \label{eq-def-chi}
  \begin{gather}
  \chi^u_{(\sfm)}=\sfU_{\{0\}}=-f,\\
  \chi_{(\sfm)}^r=r\sfR_{\{-1\}}+\sfR_{\{0\}}=\frac{r}{2}\mathscr D_aY^a+\sfR_{\{0\}},\\
  \chi_{(\sfm)}^a=\Theta^a_{\{0\}}+\frac{\Theta^a_{\{1\}}}{r}
  =-Y^a-\frac{1}{r} \left( Z^a-\sD^a f \right).\label{eq-def-chi-a}
  \end{gather}
\end{subequations}
Although the form of $\chi^\mu_{(\sfm)}$ was obtained under the good gauge, it actually does not change even if one carries out a further gauge transformation \eqref{eq-gtf-h}.
It is easy to recognize that $\chi^\mu_{(\sfm)}$ is similar to the BMS generator in GR, for example, $\xi^\mu_\text{BMS}$ in Eq.~(3.10) in Ref.~\cite{Blanchet:2020ngx}, modulo the sign difference.
They differ mainly in the subleading terms of $\chi_{(\sfm)}^r$ and $\chi_{(\sfm)}^a$, i.e., $\sfR_{\{0\}}$ and $Z^a$.
At the moment, $\sfR_{\{0\}}$ has not yet been determined, but as elucidated later, $\sfR_{\{0\}}$ is a function of $T$, $Y^a$, $Z^a$ and field perturbations. 
One shall redefine $\sfR_{\{0\}}$ such that its dependency on the field perturbations is moved to $\sfR'$ introduced in Eq.~\eqref{eq-def-exp-r}, resulting in
\begin{equation}
  \label{eq-final-r0}
  \sfR_{\{0\}}=\frac{1}{2}\left(\sD_aZ^a-\sD^2f\right),
\end{equation}
and thus, $\chi^\mu_{(\sfm)}$ is a function of $T$, $Y^a$ and $Z^a$, only.
Obviously, $\chi_{(\sfm)}^\mu$ would define the asymptotic symmetry for the mode $\sfm$, like $\xi^\mu_\text{BMS}$ in Ref.~\cite{Blanchet:2020ngx}.
By analogue, one knows that $T$ and $Y^a$ generate the supertranslation and super-Lorentz transformation, respectively.
So the asymptotic symmetry includes the familiar $\overline{\text{BMS}}$ symmetry.
In addition, the existence of $Z^a$ suggests that there is the subleading $\ebms$ symmetry parameterized by it.
In GR, BD, and dCS, $Z^a$ is zero in the Newman-Unti gauge or Bondi gauge \cite{Barnich:2010eb,Flanagan:2015pxa,Hou:2020tnd,Hou:2021oxe}.
One may want to set it to zero in Einstein-\ae ther theory, as a gauge fixing condition, then, the asymptotic symmetry group reduces to $\ebms$.
However, the analysis on the vector memory effect in Section~\ref{sec-vec-mm} suggests to keep it free, so that the treatment for all modes would be uniform.
Of course, ignoring $Z^a$, the expressions for $\chi_{(\sfm)}^\mu$ still differ from the standard ones, e.g., Eq.~(2.16) in Ref.~\cite{Flanagan:2015pxa}, by higher order terms in $1/r$.
This is due to the fact that the analysis is at the linear order in field perturbations.
Once the nonlinear analysis is done, one expects that $\chi_{(\sfm)}^\mu$ takes exactly the same form as Eq.~(2.16) in Ref.~\cite{Flanagan:2015pxa}, when $Z^a$ is set to zero.
So the asymptotic symmetry is generated by $T$, $Y^a$ and $Z^a$, and includes $\ebms$, as the leading part, and the subleading $\ebms$ symmetries.

In fact, the Minkowski spacetime also enjoys the asymptotic symmetry, as it is asymptotically flat, too.
The generator of the asymptotic symmetry in the flat spacetime is still $\chi^\mu_{(\sfm)}$, defined by Eq.~\eqref{eq-def-chi} with $\sfR_{\{0\}}$ given by Eq.~\eqref{eq-final-r0}.
It is now interesting to examine how $\chi^\mu_{(\sfm)}$ transforms the \ae ther vev, $\underline{\tilde\mfu}'^\mu=\delta^\mu_u\rightarrow\delta^\mu_u+\pd_u\chi^\mu_{(\sfm)}$.
Since $\chi^\mu_{(\sfm)}$ depends on $u$ in general, the \ae ther vev is not invariant.
Usually, in the treatment of the spontaneous symmetry breaking \cite{Srednicki:2007qs,Schwartz:2013pla}, one would like to fix vev.
So if one  followed common practice in quantum field theory, one would require $\chi^\mu_{(\sfm)}$ be independent of $u$, which means that 
\begin{equation}
  \label{eq-unitary}
  \mathscr D_aY^a=0,
\end{equation}
which amounts to the statement that the super-boost transformations are removed from the asymptotic symmetry group of Einstein-\ae ther theory.
However, since the super-boost is deeply related to the CM memory effect in GR, BD and dCS, one may not want to fix the \ae ther vev so that the intimate relation between the super-boost and the CM memory remains in Einstein-\ae ther theory.

Now, one should also check the components of the \ae ther field, as the asymptotic symmetry also transforms the \ae ther field components.
One can show that
\begin{subequations}
  \begin{gather}
    \tilde{\mathfrak u}'^u_{(\sfm)}=1+\pd_{\bar t}\sfU+\mfv''^{\bar t}-n_j\mfv''^j,\\
    \tilde{\mathfrak u}'^r_{(\sfm)}=\pd_{\bar t}\sfR+n_j\mfv''^j,\\
    \tilde{\mfu}'^a_{(\sfm)}=\pd_{\bar t}\Theta^a+\frac{e^a_j}{r}\mfv''^j,
  \end{gather}
\end{subequations}
using the coordinate transformation law of a vector field.
Since every term on the right-hand sides are series expansions in $1/r$, the components of the \ae ther field in the new coordinates have no logarithmic terms.

Up to now, one may notice that  $\sfU$, $\sfR$, and $\Theta^a$ are fixed only at the leading orders in $1/r$.
Their higher order expansion coefficients are still unknown.
This is because here, we have imposed very weak conditions on the metric components, given by Eq.~\eqref{eq-bdyc}.
These are insufficient to determine all expansion coefficients of $\sfU$, $\sfR$, and $\Theta^a$.
If it is desired, one may further require  $\tdg'^{(\sfm)}_{rr}$, $\tdg'^{(\sfm)}_{ur}+1$ and $\tdg'^{(\sfm)}_{ra}$ be zero at the orders higher than $1/r$, and this completely fixes $\sfU$, $\sfR$ and $\Theta^a$, respectively.
These requirements are weaker than, but resemble Eq.~\eqref{eq-def-nu-up}.
In the following sections, we will explicitly determine the coordinate transformations, and calculate the unphysical metric and \ae ther fields in the pseudo-Newman-Unti coordinate systems for all radiative modes.

\subsection{Memory effects}
\label{sec-mm-1}

Once one obtains the pseudo-Newman-Unti coordinates, one would like to calculate the geodesic deviation equation \eqref{eq-gdv} and inspect the memory effect due to the mode $\sfm$.
It is better to reexpress Eq.~\eqref{eq-gdv} in the pseudo-Newman-Unti coordinate system. 
Since the right-hand side of this equation involves $R_{tjtk}$, which is already of the first order in the field perturbations, let us consider the leading order part of the physical metric in the pseudo-Newman-Unti coordinates,
\begin{equation}
  \label{eq-lo-eta-pnu}
  \begin{split}
  \tilde\eta_{\mu\nu}\ud\tilde x^\mu\ud\tilde x^\nu=&-s_{\sfm}^{-2}\ud\tdu^2-2s_\sfm^{-2}\ud\tdu\ud\tdr\\
  &+(1-s_\sfm^{-2})\ud\tdr^2
  +\tdr^2\tdgamma_{ab}\ud\tdth^a\ud\tdth^b.
  \end{split}
\end{equation}
Therefore, one can identify an orthonormal basis,
\begin{equation}
  \label{eq-def-ob}
  \tilde e_{\hat0}=s_\sfm\tpd_u,\quad \tilde e_{\hat r}=-\tpd_u+\tpd_r,\quad \tilde e_{\hat a}=\tdr \tilde e_a^j\pd_j.
\end{equation}
The dual basis is
\begin{equation}
  \label{eq-def-dob}
  \tilde e^{\hat 0}=s_\sfm^{-1}(\ud\tdu+\ud\tdr), \quad \tilde e^{\hat r}=\ud\tdr,\quad \tilde e^{\hat a}=\frac{\tilde e^a_j}{\tdr}\ud x^j.
\end{equation}
In fact, near the infinity, the 4-velocities $\pd_\tau$ of the test particles approach $\tilde e_{\hat0}$, and it is natural to decompose the deviation vector $\vec S$ in the following way,
\begin{equation}
  \label{eq-s-dec}
  \vec S=\mathsf S \tilde e_{\hat r}+S^{\hat a}\tilde e_{\hat a}.
\end{equation}
Then, Eq.~\eqref{eq-gdv} can be rewritten as
\begin{subequations}
  \label{eq-gdv-pnu}
  \begin{gather}
    \frac{\ud^2\mathsf S}{\ud\tau^2}=-s^2_\sfm(\tilde R_{urur}\mathsf S+\tilde R_{uru\hat a}S^{\hat a}),\label{eq-gdv-1}\\
    \frac{\ud^2S_{\hat a}}{\ud\tau^2}=-s_\sfm^2(\tilde R_{uru\hat a}\mathsf S+\tilde R_{u\hat au\hat b}S^{\hat b}).\label{eq-gdv-2}
  \end{gather}
\end{subequations}
Here, the Riemann tensor components can be calculated using
\begin{gather}
  \label{eq-rie-c}
  \tilde R_{urur}=s_\sfm^{-2}n^jn^kR_{tjtk},\\
  \tilde R_{uru\hat a}=s_m^{-2}\tilde e^j_{\hat a}n^kR_{tjtk},\\
  \tilde R_{u\hat au\hat b}=s_\sfm^{-2}\tilde e_{\hat a}^j\tilde e_{\hat b}^kR_{tjtk}.
\end{gather}
Usually, one call $\mathsf S$ the longitudinal mode, and $S^{\hat a}$ the transverse modes.

Suppose the radiative mode $\sfm$ exists at $\tdr\rightarrow\infty$ from the time $\tdu_0$ to $\tdu_f$.
Integrating Eq.\eqref{eq-gdv-pnu} gives the total changes $\Delta\mathsf S$ and $\Delta S^{\hat a}$ between $\tdu_0$ and $\tdu_f$.
If $\Delta\mathsf S\ne0$ or $\Delta S^{\hat a}\ne0$, one claims that there exists the memory effect.
By studying the functional dependencies of $\Delta\mathsf S$ and $\Delta S^{\hat a}$ on the radiative modes, one can decipher the relation between memories and asymptotic symmetries, as explicitly demonstrated in Sections~\ref{sec-ten-mm}, \ref{sec-vec-mm} and \ref{sec-sca-mm}.

\section{Pseudo-Newman-Unti coordinates for tensor modes}
\label{sec-pnu-t}

In this section, the pseudo-Newman-Unti coordinate system will be determined for the tensor mode. 
For this purpose, one should solve Eq.~\eqref{ettjk} and write $h_{jk}^\text{TT}$ using multipolar moments in the physical frame.
This would be done in Section~\ref{sec-ten-vac-sols}. 
Then, work in the unphysical frame with $\sigma=s_\sfg^2$.
Now, fix the gauge, and perform a suitable infinitesimal transformation Eq.~\eqref{eq-gtf-h} such that $\bar h''_{\mu\nu}$ takes a simple form, in Section~\ref{sec-t-gf}.
In Section~\ref{sec-psu-t-c}, one gets the transformation Eq.~\eqref{eq-def-ctf-2-nu}, and the pseudo-Newman-Unti coordinates are obtained, together with the asymptotic symmetries.
The unphysical metric and \ae ther fields will also be explicitly calculated.
Finally, the memory effects of the tensor mode are discussed in Section~\ref{sec-ten-mm}.

\subsection{Vacuum multipolar solution}
\label{sec-ten-vac-sols}


Consider the tensor equation \eqref{ettjk} first.
In the vacuum,  one knows that the solution is \cite{Thorne:1980ru,Blanchet:1985sp}
\begin{equation*}
  \label{eq-htt-sol-g}
  \begin{split}
    h_{jk}^\text{TT}=&\sum_{l=0} \left[ \pd_{jkL} \left( \frac{E_L}{r} \right)+\delta_{jk}\pd_L \left( \frac{F_L}{r} \right) \right]+\\
    &\sum_{l=1}\pd_{L-1} \left[ \pd_{(j}\left(\frac{G_{k)L-1}}{r}\right) +\epsilon_{pq(j}\pd_{k)p} \left( \frac{H_{qL-1}}{r} \right) \right]\\
    &+\sum_{l=2}\pd_{L-2} \left[  \left(\frac{I_{jkL-2}}{r}\right) + \pd_{p} \left( \frac{\epsilon_{pq(j}J_{k)qL-2}}{r} \right) \right],
  \end{split}
\end{equation*}
in general.
Here, the symmetric-tracefree tensors $(E_L,F_L,G_L,H_L,I_L,J_L)$ are all functions of $s_{\mathsf g}t-r$, the retarded time associated with the tensor GW.
What essentially differentiates Eq.~\eqref{ettjk} from Eq.~(2.2) in Ref.~\cite{Blanchet:1985sp} is that $h_{jk}^{\text{TT}}$ here satisfies the transverse-tracefree condition, while $h_{\mu\nu}$ there is merely transverse.
Then, imposing the transverse-tracefree condition ($\pd^kh_{jk}^\text{TT}=0$ and $\delta^{jk}h_{jk}^\text{TT}=0$) leads to 
\begin{subequations}
\begin{gather}
  F=E^{(2)}=0, \quad F_L=\sinv{{\mathsf g}}{2}E_L^{(2)}\;(l\ge1), \label{eq-fe-0}\\
  G_j^{(2)}=0,\quad G_L=-4\sinv{{\mathsf g}}{2}E_L^{(2)}\;(l\ge1),\label{eq-gj2-0}\\
  I_L=2\sinv{{\mathsf g}}{4}E_L^{(4)}\;(l\ge2),\\
  H_j^{(2)}=0,\quad J_L=-\sinv{{\mathsf g}}{2}H_L^{(2)}\;(l\ge2),\label{eq-hj2-0}
\end{gather}
\end{subequations}
where the superscript $(n)$ indicates the $n$-th order partial derivative with respect to $t$.
The metric perturbation is now given by 
\begin{equation}
  \label{eq-htt-s-1}
  \begin{split}
    &h_{jk}^\text{TT}=\sum_{l=0} \pd_{jkL}\frac{E_L}{r}+\delta_{jk}\sum_{l=1}\pd_L\frac{E^{(2)}_L}{s_{\mathsf g}^2r}\\
    &-4\sum_{l=1}\pd_{L-1(j}\frac{E_{k)L-1}^{(2)}}{s_{\mathsf g}^2r}+\sum_{l=1}\epsilon_{pq(j}\pd_{k)pL-1}\frac{H_{qL-1}}{r}\\
    &+\sum_{l=2} \pd_{L-2}\left(2\frac{E_{jkL-2}^{(4)}}{s_{\mathsf g}^4r}-\pd_{p}\frac{\epsilon_{pq(j}H_{k)qL-2}^{(2)}}{s_{\mathsf g}^2r}  \right).
  \end{split}
\end{equation}
From this expression, one notices that the metric perturbation is determined by two sets of transverse-tracefree tensors, $E_L$ and $H_L$, which is consistent with the fact that there are two tensorial degrees of freedom.

\subsection{Gauge fixing}
\label{sec-t-gf}

According to the discussion in Section~\ref{sec-sch}, one should now fix the gauge such that $\varepsilon_j=0$ and $\omega=\gamma=0$ \cite{Foster:2006az}.
If one considers only the tensor mode, then one finds out that
\begin{equation}
  h_{jk}=h_{jk}^\text{TT},
\end{equation}
and the remaining components of $h_{\mu\nu}$ vanish, so do all of the components of $\mfv^\mu$.
Now, one should work in the unphysical frame with $\sigma=s_{\mathsf g}^2$, so that the tensor mode travels at the speed of one, as measured by the unphysical metric $g'^{(\sfg)}_{\mu\nu}$.
Then, perform the coordinate transformation $t\rightarrow \bar t=s_\sfm t,\,x^j\rightarrow x^j$.
So, the unphysical metric and \ae ther fields are
\begin{subequations}
\begin{gather}
  g'^{(\sfg)}_{\bar t\bar t}=-1,\quad
  g'^{(\sfg)}_{\bar tj}=0,\quad
  g'^{(\sfg)}_{jk}
  =\delta_{jk}+h_{jk}^\text{TT},\\
  \mfu'^{\bar t}_{(\sfg)}=1,\quad
  \mfu'^j_{(\sfg)}=0.
\end{gather}
\end{subequations}
This way, the leading order parts of the metric and \ae ther fields take the usual forms.
The trace-reversed metric perturbation $\bar h'_{\mu\nu}$ is
\begin{equation}
  \label{eq-hbar-o-g}
  \bar h'_{\bar t\bar t}=0,\quad
  \bar h'_{\bar tj}=0,\quad
  \bar h'_{jk}=h_{jk}^\text{TT}.
\end{equation}
In this gauge, $\bar h'_{\mu\nu}$, given by Eqs.~\eqref{eq-hbar-o-g} and \eqref{eq-htt-s-1}, is drastically different from the one in GR, referring to Eq.~(2.3) in Ref.~\cite{Blanchet:2020ngx}.

Now, one tries to make $\bar h'_{\mu\nu}$ look like the one in GR  as much as possible.
This means one should perform the gauge transformation parameterized by Eq.~\eqref{eq-def-inf-gtf}, with $\sW_L$, $\sV_L$, $\sT_L$, and $\sR_L$ all functions of $s_\sfg t-r$.
After some trial and error, one takes the following good gauge conditions,
\begin{equation*}
  \sW_L= \frac{E_L^{[1]}}{2},\quad 
  \sV_L=\frac{E_L}{2},\quad
  \sT_L=-2E_L^{[2]},\quad \sR_L=\frac{H_L}{2},
\end{equation*}
where $[n]$ means to take the $n$-th order derivative with respect to $u$.
Furthermore, if one defines
\begin{gather}
  \label{eq-ids-gr}
  E_L^{[2]}=2\frac{(-1)^l}{l!}M_L \;(l\ge0),\\ H_L^{[1]}=-8\frac{(-1)^l}{l!}\frac{l}{l+1}S_L\;(l\ge1),
\end{gather}
then, one has 
\begin{subequations}
  \label{eq-hbar-ttcs}
\begin{gather}
  \bar h''_{\bar t\bar t}= 4 \sum_{l=0}\frac{(-1)^l}{l!}\pd_L\frac{M_L}{r},\\
  \bar h''_{\bar tj}= 4\sum_{l=1} \frac{(-1)^l}{l!}\pd_{L-1}\left( \frac{M_{jL-1}^{[1]}}{r} + \frac{l\epsilon_{jpq}}{l+1}\pd_{p}\frac{S_{qL-1}}{r} \right),\\
  \begin{split}
  \bar h''_{jk}= 4\sum_{l=2}\frac{(-1)^l}{l!}\pd_{L-2} &\Bigg(  \frac{M_{jkL-2}^{[2]}}{r}\\
  &+\frac{2l}{l+1}\pd_{p}\frac{\epsilon_{pq(j}S_{k)qL-2}^{[1]}}{r} \Bigg).
  \end{split}
\end{gather}
These equations are equivalent to Eq.~(2.3) in Ref.~\cite{Blanchet:2020ngx}. 
$M_L$ and $S_L$ are the mass and current multipole moments, respectively.
Also, Eqs.~\eqref{eq-gj2-0} and \eqref{eq-hj2-0} implies that the linear momentum $P_j\equiv M_j^{[1]}$ and angular momentum $S_j$ are constant as in GR.
Note that $M=E^{[2]}/2=0$ according to Eq.~\eqref{eq-fe-0}, so the mass monopole is zero in the tensor sector in Einstein-\ae ther theory.
Despite this minor difference, these expressions suggest that one can repeat the calculation in Ref.~\cite{Blanchet:2020ngx} to obtain the Newman-Unti coordinates for the tensor mode, and analyze the asymptotic behaviors.
However, in this work we will not carry out this computation, as it is impossible to find the well-behaved Newman-Unti coordinates for the vector and scalar modes using the same method as discussed in Section~\ref{sec-sch}.

One should also calculate the perturbation of the \ae{}ther field, given by,
\begin{gather}
  \mfv''^{\bar t}=-\sum_{l=0}\frac{(-1)^l}{l!}\pd_L\frac{M_L}{r},\\
  \begin{split}
  \mfv''^j=&\sum_{l=0}\frac{(-1)^l}{l!}\pd_{jL}\frac{\hat M_L}{r}-4\sum_{l=1}\frac{(-1)^l}{l!}\\
  & \times\pd_{L-1}\Bigg( \frac{M_{jL-1}^{[1]}}{r}
  +\frac{l}{l+1}\epsilon_{jpq}\pd_{p}\frac{S_{qL-1}}{r} \Bigg),
  \end{split}
\end{gather}
\end{subequations}
where $\hat M_L=\int M_L\ud\bar t$.
These equations are presented for completeness.

\subsection{Constructing pseudo-Newman-Unti coordinates}
\label{sec-psu-t-c}

In the following, we will explicitly obtain $\sfU$, $\sfR$ and $\Theta^a$ in Eq.~\eqref{eq-def-ctf-2-nu} based on the discussion in Section~\ref{sec-sch}.
The metric components and the \ae ther field would also be calculated.

Let us consider $\chi^\mu_{(\sfg)}$ first, defined in Eq.~\eqref{eq-def-chi}, which includes the leading order parts of $\sfU$, $\sfR$ and $\Theta^a$. 
According to the discussion in Section~\ref{sec-sch}, $\chi_{(\sfg)}^\mu$ can be determined by requiring the components of $\tilde g'^{(\sfg)}_{\mu\nu}$ in the pseudo-Newman-Unti coordinates to obey certain boundary conditions \eqref{eq-bdyc} near the infinity $r\rightarrow\infty$.
The boundary conditions imposed in that section are relatively weak for the tensor mode.
As one can check, it is possible to further impose $\tdg'^{(\sfg)}_{ra}\sim\order{1/r}$, compared with Eq.~\eqref{eq-bdyc-1-3}.
This results in $Z^a=-4e^a_jP_j$, and thus $\sfR_{\{0\}}=5n_jP_j-\mathscr D^2f/2$, by requiring $\det(\tdg'^{(\sfg)}_{ab})$ vanish at $\order{r^3}$.
However, this stronger condition may not be consistently imposed in the vector sector.
In order to make the treatment uniform, we will stick with conditions \eqref{eq-bdyc} for the tensor mode.
Meaningly, one should let $Z^a$ be free, and that leads to
\begin{equation}
  \label{eq-def-r0-ten}
  \sfR_{\{0\}}=\frac{1}{2}\left(\tdcd_aZ^a-\tdcd^2f\right).
\end{equation}
Therefore, $\chi^\mu_{(\sfg)}$ is completely parameterized by $T$, $Y^a$ and $Z^a$, and the asymptotic symmetry includes the $\ebms$ and the subleading $\ebms$ symmetries as discussed in the previous section.
This is different from GR \cite{Blanchet:2020ngx}, BD \cite{Hou:2020tnd,Tahura:2020vsa}, and dCS \cite{Hou:2021oxe,Hou:2021bxz}.

Now, one can determine the higher order terms of $\sfU$, $\sfR$, and $\Theta^a$. 
It turns out that
\begin{subequations}
  \label{eq-ctf-ten}
  \begin{gather}
    \begin{split}
    \sfU=&\chi_{(\sfg)}^u+4\sum_{l=1}\frac{1}{l!}\sum_{k=1}^l\frac{2k-1}{(l+k)(l+k-1)}\\
    &\times a_{kl}\frac{n_LM_L^{[l-k]}}{r^k},
    \end{split}\\
    \begin{split}
    \sfR=&\chi_{(\sfg)}^r+2\sum_{l=2}\frac{1}{l!}\sum_{k=1}\frac{(l-k)(l+3k-1)}{(l+k)(l+k-1)(k+1)}\\
    &\times a_{kl}\frac{n_LM_L^{[l-k]}}{r^k},
    \end{split}\\
    \begin{split}
    \Theta^a=&\chi_{(\sfg)}^a-4\frac{e^a_j}{r}\sum_{l=1}\frac{1}{l!}\sum_{k=1}^l\frac{a_{kl}}{(l+k)(k+1)}\frac{n_{L-1}}{r^k} \\
    &\times\left( \frac{2k^2-l}{l+k-1}M_{jL-1}^{[l-k]}+\frac{2kl}{l+1}\epsilon_{jpq}n_pS_{qL-1}^{[l-k]} \right),\label{eq-ctf-ten-th}
    \end{split}
  \end{gather}
\end{subequations}
  where 
  \begin{equation}
    \label{eq-def-akl}
    a_{kl}=\frac{(l+k)!}{2^kk!(l-k)!}.
  \end{equation}
One can compare these equations with Eq.~(3.2) in Ref.~\cite{Blanchet:2020ngx}. 
First, let us compare the leading order terms represented by $\chi_{(\sfg)}^\mu$ with $\xi^\mu_\text{BMS}$ given by Eq.~(3.10) in Ref.~\cite{Blanchet:2020ngx}.
Although at the subleading orders $\chi_{(\sfg)}^\mu$ is different from $\xi^\mu_\text{BMS}$ by $\sfR_{\{0\}}$ and $Z^a$, their leading order parts are actually the same, modulo the difference in the sign convention.
Second, let us focus on the higher order terms in Eq.~\eqref{eq-ctf-ten}, and compare them with the corresponding terms in Eq.~(3.2) in Ref.~\cite{Blanchet:2020ngx}.
It is easy to find that these two sets of equations share the same parts that are proportional to the symbols $a_{kl}$.
The remaining parts are different.
That is, here, there are no $\ln r$-terms, as we have imposed weaker conditions on the metric components.

In the pseudo-Newman-Unti coordinates, some of the metric components are thus,
\begin{gather}
  \label{eq-tg-rr-ra-ur-ten}
  \tdg'^{(\sfg)}_{rr}=-\frac{4\tdn_jP_j}{\tdr},\quad \tdg'^{(\sfg)}_{ur}=-1,\quad \tdg'^{(\sfg)}_{ra}=-Z_a,
\end{gather}
by Eq.~\eqref{eq-gc-pnu}.
As one can see, $\tdg'^{(\sfg)}_{rr}$ is not zero at a finite $\tdr$, and $\tdg'^{(\sfg)}_{ra}$ can be freely specified.
So the pseudo-Newman-Unti coordinate system for the tensor mode is almost in the Newman-Unti gauge.
If one works in the center-of-mass frame with $P_j=0$, and chooses a gauge such that $Z_a=0$, the Newman-Unti gauge is recovered.
The remaining metric components are
\begin{subequations}
  \label{eq-tdg-uab}
\begin{gather}
  \tdg'^{(\sfg)}_{uu}=-1-(\tdcd^2+2)f^{[1]}+\frac{2}{\tdr} \left( m+\sum_{k=1}\frac{K_k}{\tdr^k} \right),\\
  \begin{split}
  \tdg'^{(\sfg)}_{ua}=&\frac{1}{2}\tdcd^bc_{ab}+\frac{1}{\tdr} \left( \frac{2}{3}N_a+\tilde e^j_a\sum_{k=1}\frac{P^j_k}{\tdr^k} \right)\\
  &+\Delta\tdg'^{(\sfg)}_{ua},\label{eq-tdg-ua-ten}
  \end{split}\\
  \begin{split}
  \tdg'^{(\sfg)}_{ab}=&\tdr^2 \left[\tdgamma_{ab}+2\tdcd_{\langle a}Y_{b\rangle}\right.\\
  &\left.+\frac{1}{\tdr} \left( c_{ab}+\tilde e^j_{\langle a}\tilde e^k_{b\rangle}\sum_{l=1}\frac{Q^{jk}_l}{\tdr^l} \right)\right]
  +\Delta\tdg'^{(\sfg)}_{ab},\label{eq-tdg-ab-ten}
  \end{split}
\end{gather}
where $Y_a=\tdgamma_{ab}Y^b$, $\Delta\tdg'^{(\sfg)}_{ua}=\tdcd_a\tdcd_bZ^b/2$, and $\Delta\tdg'^{(\sfg)}_{ab}=2\tdr\tdcd_{\langle a}Z_{b\rangle}$ with $Z_a=\tdgamma_{ab}Z^b$.
\end{subequations}
Note that $\Delta\tdg'^{(\sfg)}_{ab}$ is traceless, like $c_{ab}$.
Here, these metric components are put in such a form that the so-called Bondi data can be easily read off.
Indeed, $m$ and $N_a$ resemble the Bondi mass aspect and angular momentum aspect \cite{Barnich:2010eb,Madler:2016xju,Blanchet:2020ngx}, 
\begin{subequations}
  \begin{gather}
    m=\sum_{l=0}\frac{(l+1)(l+2)}{2l!}\tdn_LM_L^{[l]},\label{eq-def-mas}\\
    N_a=\tilde e_a^j\sum_{l=1}\frac{(l+1)(l+2)}{2(l-1)!}\tdn_{L-1} \bigg( M_{jL-1}^{[l-1]}\nonumber\\
    +\frac{2l}{l+1}\tilde\epsilon_{jpq}\tdn_pS_{qL-1}^{[l-1]} \bigg),\label{eq-def-angas}
  \end{gather}
\end{subequations}
at the linear order in the field perturbations.
The tensor $c_{ab}$ looks like the shear tensor, given by
\begin{equation}
  \label{eq-def-shear-2}
  \begin{split}
  c_{ab}=&4\tilde e^j_{\langle a}\tilde e^k_{b\rangle}\sum_{l=2}\frac{\tdn_{L-2}}{l!}\left( M_{jkL-2}^{[l]}\right.\\
  &\left.-\frac{2l}{l+1}\tilde\epsilon_{jpq}\tdn_pS_{kqL-2}^{[l]} \right)-2\tdcd_{\langle a}\tdcd_{b\rangle}f.
  \end{split}
\end{equation}
And the remaining symbols $K_k$, $P^j_k$, and $Q^{ij}_k$ are 
\begin{subequations}
  \begin{gather}
    K_k=\frac{1}{(k+1)(k+2)}\sum_{l=k}\frac{(l+1)(l+2)}{l!}a_{kl}\tdn_LM_L^{[l-k]},\\
    P^j_k=\frac{2}{k+3}\sum_{l=k+1}\frac{l+2}{l!}a_{k+1,l}\tdn_{L-1} \Big( M_{jL-1}^{[l-k-1]}\nonumber\\
    +\frac{2l}{l+1}\tilde\epsilon_{jpq}\tdn_pS_{qL-1}^{[l-k-1]} \Big),\\
    Q^{ij}_k=4\frac{k-1}{k+1}\sum_{l=k}\frac{a_{kl}}{l!}\tdn_{L-2} \Big( M_{ijL-2}^{[l-k]}\nonumber\\
    +\frac{2l}{l+1}\tilde\epsilon_{ipq}\tdn_pS_{jqL-2}^{[l-k]} \Big).
  \end{gather}
\end{subequations}
Now, let us compare Eq.~\eqref{eq-tdg-uab} with Eq.~(3.14) in Ref.~\cite{Blanchet:2020ngx}.
These two sets of equations are basically the same, except that $\tdg'^{(\sfg)}_{ua}$ and $\tdg'^{(\sfg)}_{ab}$ both have extra terms, i.e., $\Delta\tdg'^{(\sfg)}_{ua}$ and $\Delta\tdg'^{(\sfg)}_{ab}$, respectively.
These extra terms and $\tdg'^{(g)}_{ra}$ are due to the presence of $Z_a$, and can be viewed as pure gauge in the tensor sector.

The form of $c_{ab}$ as defined by Eq.~\eqref{eq-def-shear-2} indicates its transformation under the supertranslation
 $\tdu\rightarrow \tdu+\alpha(\tdth^a)$, i.e., 
\begin{equation}
  \label{eq-stf-cab}
c_{ab}\,\rightarrow\,  c'_{ab}=c_{ab}-2\tdcd_{\langle a}\tdcd_{b\rangle}\alpha.
\end{equation}
This is similar to the familiar transformation rule of the shear tensor under the supertranslation in GR, BD and dCS \cite{Barnich:2010eb,Flanagan:2015pxa,Hou:2020tnd,Tahura:2020vsa,Hou:2021oxe}.
This rule will be useful for interpreting the tensor displacement memory effect.
Under the transformation described by $Y^a$, 
\begin{equation}
  \label{eq-stf-cab-y}
  c_{ab}\,\rightarrow\, c_{ab}-2u\tdcd_{\langle a}\tdcd_{b\rangle}\tdcd_aY^a.
\end{equation}
Of course, neither of Eqs.~\eqref{eq-stf-cab} and \eqref{eq-stf-cab-y} is exactly the same as those appearing in GR, BD and dCS \cite{Barnich:2010eb,Flanagan:2015pxa,Hou:2020tnd,Tahura:2020vsa,Hou:2021oxe}, as several terms are missing.
However, these missing terms are products between $\alpha$, $Y^a$ or their derivatives and $c_{ab}$ or its derivatives, so they are of the second order in the field perturbations, and can be ignored in the linearized analysis.
Under $Z^a$-transformation, $c_{ab}$ is invariant.

Finally, the \ae ther field has the following components 
\begin{subequations}
  \label{eq-ae-t}
 \begin{gather}
  \begin{split}
  \tilde{\mathfrak u}'^u_{(\sfg)}
  =&1-\frac{\tdcd_a Y^a}{2}-2\sum_{l=1}\frac{1}{l!}\sum_{k=0}^l \alpha^u_{kl}a_{kl}\frac{\tdn_LM_L^{[l-k]}}{\tdr^{k+1}}\\
  &+\sum_{l=0}\frac{l+1}{l!}a_{ll}\frac{\tdn_L\hat M_L}{\tdr^{l+2}},
  \end{split}\\
  \begin{split}
    \label{eq-mfu-r-nu}
    \tilde{\mathfrak u}'^r_{(\sfg)}=&-\frac{\tdcd^2\tdcd_a Y^a}{4}-\sum_{l=2}\frac{1}{l!}\sum_{k=0}^l\alpha^r_{kl} a_{kl}\frac{\tdn_LM_L^{[l-k]}}{\tdr^{k+1}}\\
    &-\sum_{l=0}\frac{l+1}{l!}a_{ll}\frac{\tdn_L\hat M_L}{\tdr^{l+2}},
  \end{split}\\
  \begin{split}
    \tilde{\mfu}'^a_{(\sfg)}=&\frac{\tilde{\mathscr D}^a\tdcd_bY^b}{2\tdr}+\frac{\tilde e^a_j}{\tdr} \sum_{l=1}\tdn_{L-1}\left[\frac{a_{ll}}{(l-1)!}\frac{\hat M_{jL-1}}{\tdr^{l+2}}\right.\\
    &+2\frac{1}{l!}\sum_{k=0}^l \alpha^a_{kl}a_{kl}\frac{M_{jL-1}^{[l-k]}}{\tdr^{k+1}}+4\frac{l}{(l+1)!}\\
    &\left.\times\sum_{k=0}^l\frac{l-2k-2}{k+2}a_{kl}\tilde\epsilon_{jpq}\frac{\tdn_{p}S_{qL-1}^{[l-k]}}{\tdr^{k+1}} \right],
  \end{split}
\end{gather}
\end{subequations}
where there are several complicated factors given by
\begin{subequations}
\begin{gather}
  \alpha_{kl}^u=\frac{l-k}{l+k}\frac{4k+3}{k+1}-\frac{k^2}{(l+k)(l-k+1)},\\
  \begin{split}
  \alpha^r_{kl}=&\frac{(l-k)(l-k-1)(l+3k+4)}{(l+k)(k+2)(k+1)}\\
  &-\frac{3l^2-(8k-3)l+k(3k-5)}{(l+k)(l-k+1)},
  \end{split}\\
  \begin{split}
  \alpha^a_{kl}=&\frac{kl}{(l+k)(l-k+1)}\\
  &-\frac{l-k}{l+k}\frac{l-2(k+1)(2k+3)}{(k+2)(k+1)}.
  \end{split}
\end{gather}
\end{subequations}
Unlike the metric components in Eqs.~\eqref{eq-tg-rr-ra-ur-ten} and \eqref{eq-tdg-uab}, the \ae ther field is expressed directly in terms of the mass and current multipole moments $M_L$ and $S_L$, as it is not easy to recognize the physical meaning of their combinations.

As one can see, Eq.~\eqref{eq-ae-t} is very complicated, and it is difficult to interpret the physical meaning of the terms. 
Similar situation will happen to the metric and \ae ther fields for the vector and scalar modes. 
Therefore, in Sections~\ref{sec-pnu-v} and \ref{sec-pnu-s}, we will not present the complete expressions for the metric and \ae ther fields.
Only leading order terms in $1/\tdr$ will be given.

\subsection{Tensor memory effects}
\label{sec-ten-mm}

Let us now examine the memory effects associated with the tensor modes. 
One can show that at $\order{1/\tdr}$, $\tilde R_{urur}=\tilde R_{uru\hat a}=0$,
then, simply consider,
\begin{equation}
  \label{eq-gdv-t}
  \frac{\ud^2S_{\hat a}}{\ud \tau^2}=-s_\sfg^2\tilde R_{u\hat au\hat b}S^{\hat b}=\frac{1}{2\tdr}s_\sfg^2c^{[2]}_{\hat a\hat b}S^{\hat b}+\order{\frac{1}{\tdr^2}}.
\end{equation}
Provided that the initial relative velocity of the test particles is nonzero, one integrates this equation twice to get
\begin{equation}
  \label{eq-def-tmm}
  \begin{split}
  \Delta S_{\hat a}\approx &\left.S^{[1]}_{\hat a}\right|_0\Delta\tilde u+\frac{\Delta c_{\hat a\hat b}}{2\tdr}\left.S^{\hat b}\right|_0\\
  &+\frac{1}{\tdr} \left[ \frac{c_{\hat a\hat b}(\tilde u_f)+c_{\hat a\hat b}(\tilde u_0)}{2}\Delta\tdu-\Delta\mathcal C_{\hat a\hat b} \right]\left.S^{[1]}_{\hat a}\right|_0,
  \end{split}
\end{equation}
where $S_{\hat a}|_0$ and $\left.S^{[1]}_{\hat a}\right|_0$ are the initial relative displacement and velocity at the time $\tdu_0$, when the GW arrives, $\Delta \tdu=\tdu_f-\tdu_0$ with $\tdu_f$ the time when the GW disappears,
$\Delta c_{\hat a\hat b}=c_{\hat a\hat b}(\tdu_f)-c_{\hat a\hat b}(\tdu_0)$, and $\Delta \mathcal C_{\hat a\hat b}=\int_{\tdu_0}^{\tdu_f}c_{\hat a\hat b}(\tdu')\ud\tdu'$.

If one compares this equation with, for example, Eqs.~(62) and (63) in Ref.~\cite{Hou:2021oxe}, one realizes that the term with $\Delta c_{\hat a\hat b}$ is the (leading) displacement memory effect, and the one with $\Delta\mathcal C_{\hat a\hat b}$ contains the spin and the CM memory effects. 
See also Ref.~\cite{Tahura:2020vsa}.
One can always express 
\begin{equation}
  \label{eq-shear-dec}
  c_{ab}=\tdcd_{\langle a}\tdcd_{b\rangle}\sfc+\tilde\epsilon_{c(a}\tdcd_{b)}\tdcd^c\sfd,
\end{equation}
where $\tilde\epsilon_{ab}$ is the volume element on a unit 2-sphere, and $\sfc$ and $\sfd$ are the electric and magnetic parts of $c_{ab}$, respectively.
The spin memory is given by $\int_{\tdu_0}^{\tdu_f}\sfd\ud\tdu$, while the CM memory is $\int_{\tdu_0}^{\tdu_f}\tdu\sfc^{[1]}_o\ud\tdu$.
Here, $\sfc_o$ is the part of $\sfc$ that is related to the so-called ordinary memory effect.
In fact, in the linearized theory, one can show that the evolution equation of $m$ is
\begin{equation}
  \label{eq-evo-mas}
  m^{[1]}=\frac{1}{4}\tdcd_a\tdcd_bN^{ab},
\end{equation}
with $N_{ab}=c^{[1]}_{ab}$ the news tensor.
Integrating this equation over the time $\tdu\in(\tdu_0,\tdu_f)$ gives 
\begin{equation}
  \label{eq-def-co}
  \Delta m=\frac{1}{8}\tdcd^2(\tdcd^2+2)\Delta\sfc_o.
\end{equation}
Of course, if one could find the nonlinear evolution equation for $m$, one would obtain a similar equation for the null memory part $\sfc-\sfc_o$.
Therefore, like in GR, BD and dCS, the displacement memory, spin and CM memory are related to the shear tensor $c_{ab}$. 
Of course, in Einstein-\ae ther theory, $c_{ab}$ is the shear tensor of the unphysical metric tensor $\tdg'^{(\sfg)}_{\mu\nu}$.

So if a vacuum state for the tensor mode is also the one described by $c_{ab}=\tdcd_{\langle a}\tdcd_{b\rangle}\mathsf C$ with $\mathsf C=\mathsf C(\tdth^a)$, then the displacement memory effect in Einstein-\ae ther theory is also the vacuum transition, as
\begin{equation}
  \label{eq-ten-dis}
\Delta c_{ab}=\tdcd_{\langle a}\tdcd_{b\rangle}\Delta\mathsf C,
\end{equation}
where $\Delta\mathsf C=-2\alpha$ with $\alpha$ defined in the sentence above Eq.~\eqref{eq-stf-cab}.
One can expand $\Delta\sfc$ using $\tdn_L$ in the following way \cite{Maggiore:1900zz},
\begin{equation}
  \label{eq-def-Dcexp}
  \Delta\sfc=\sum_{l=0}\tdn_L\Delta \sfc_L.
\end{equation}
Then by the definition \eqref{eq-def-shear-2}, one can also show that
\begin{equation}
  \label{eq-dis-ten-mag-l}
  \Delta\sfc_L=2\frac{(l-2)!}{(l!)^2}\Delta M_L^{[l]},
\end{equation}
for $l\ge2$, so the tensor displacement memory effect is of the so-called electric parity as in GR \cite{Bieri:2018asm}.

Because of Eq.~\eqref{eq-ten-dis}, one expects that the flux-balance law associated with the supertranslation could also be rewritten as the constraint equation on the displacement memory effect for the tensor mode as in Refs.~\cite{Flanagan:2015pxa,Hou:2020wbo,Tahura:2020vsa,Hou:2021bxz}.
Equivalently, one can integrate the evolution equation for the Bondi mass aspect to obtain the constraint.
Unfortunately, since we are working in the linear theory, 
the evolution equation for the Bondi mass aspect $m$ does not include the contribution of the energy flux of the GW, which shall be quadratic in $N_{ab}$.
Therefore, one cannot get the constraint on the displacement memory effect.
Constructing the pseudo-Newman-Unti coordinate system up to second order in the field perturbations could provide the quadratic terms  \cite{Blanchet:2020ngx}, but this is beyond the scope of the present work.
Similarly, for the spin and the CM memory effects, their constraint equations shall also rely on the construction of the evolution equation of the angular momentum aspect $N_a$ up to the second order in the field perturbations, which cannot be obtained in the linearized analysis.
So here, we will not try to calculate the constraints on memory effects.

In summary, the leading displacement memory in the tensor sector shares many characteristics with the one in GR.
They are both the variation in the shear tensor $c_{ab}$, can be viewed as the vacuum transition and are deeply related to the supertranslations.
Although not explicitly verified in the linear analysis, the triangular equivalence between the leading displacement memory, the supertranslation and the leading soft graviton theorem shall also hold in Einstein-\ae ther theory, which can be examined in the nonlinear analysis.

\section{Pseudo-Newman-Unti coordinates for vector modes}
\label{sec-pnu-v}

In this section, the pseudo-Newman-Unti coordinates will be obtained for the vector mode following the general scheme in Section~\ref{sec-sch}.
The structure is similar to Section~\ref{sec-pnu-t} for the tensor mode.

\subsection{Vacuum multipolar solution}

Now, the vector equation \eqref{eomsig} is to be solved. 
The general solution to Eq.~\eqref{eomsig} is \cite{Blanchet:1985sp,Thorne:1980ru}
\begin{equation*}
  \label{eq-sig-i}
  \Sigma_j=\sum_{l=0}\pd_{jL}\frac{B_L}{r}+\sum_{l=1} \pd_{L-1}\left( \frac{C_{jL-1}}{r}+\epsilon_{jpq}\pd_{p}\frac{D_{qL-1}}{r} \right),
\end{equation*}
for some transverse-tracefree tensors $B_L,\,C_L$, and $D_L$, all functions of the retarded time $s_{\mathsf v}t-r$.
The transverse condition leads to 
\begin{equation}
  \label{eq-tc-sig}
  \pd^j\Sigma_j=\sum_{l=0}\pd_L\frac{B_L^{(2)}}{s_{\mathsf v}^2r}+\sum_{l=1}\pd_L\frac{C_L}{r}=0,
\end{equation}
which implies that 
\begin{equation}
  B^{(2)}=0,\quad \sinv{{\mathsf v}}{2}B_L^{(2)}+C_L=0\;(l\ge1).
\end{equation}
Therefore, one has 
\begin{equation}
  \label{eq-sig-f}
  \begin{split}
  \Sigma_j= \sum_{l=0}\pd_{jL}\frac{B_L}{r}-\sum_{l=1}&\pd_{L-1}\bigg(\frac{B_{jL-1}^{(2)}}{s_{\mathsf v}^2r}\\
  &- \epsilon_{jpq}\pd_{p}\frac{D_{qL-1}}{r}\bigg).
  \end{split}
\end{equation}
Again, there are two sets of transverse-tracefree tensors, $B_L$ and $D_L$, here. 
Indeed, there are two vectorial degrees of freedom.
$\Xi_j$ can thus be obtained using Eq.~\eqref{eq-xieom}.

\subsection{Gauge fixing}

To fix the gauge, one also sets $\varepsilon_j=0$ and $\omega=\gamma=0$ as for the tensor mode. 
Then, perform the coordinate transformation $\bar t=s_\sfv t$ such that in the coordinates $(\bar t,x^j)$, $g'^{(\sfv)}_{\bar t\bar t}=-1=-\mfu'^{\bar t}_{(\sfv)}$, $g'^{(\sfv)}_{jk}=\delta_{jk}$, and
\begin{subequations}
\begin{gather}
  \begin{split}
  g'^{(\sfv)}_{\bar tj}=\sum_{l=0}\pd_{jL}\frac{\check B_L}{r}
  -\sum_{l=1} &\pd_{L-1}\bigg( \frac{\check B_{jL-1}^{[2]}}{r}\\
  & -\epsilon_{jpq}\pd_{p}\frac{\check D_{qL-1}}{r}\bigg),
  \end{split}\\
  \begin{split}
  \mfu'^j_{(\sfv)}=\sum_{l=0}\pd_{jL}\frac{\mathring B_L}{r}-\sum_{l=1}&\pd_{L-1}\bigg(\frac{\mathring B_{jL-1}^{[2]}}{r}\\
  &-\epsilon_{jpq}\pd_{p}\frac{\mathring D_{qL-1}}{r}\bigg),
  \end{split}
\end{gather}
where $[n]$ also means to take the $n$-th derivative with respect to $u$.
Here, one defines
\begin{gather}
  \check B_L=\mathsf k_1B_L,\quad
  \check D_L=\mathsf k_1D_L,\nonumber\\ \mathsf k_1=-\frac{1+c_+s_{\mathsf g}^2-s_\sfv ^2}{s_\sfv },\label{eq-def-cb}\\
  \mathring B_L=\mathsf k_2B_L,\quad
  \mathring D_L=\mathsf k_2D_L,\nonumber\\ \mathsf k_2=\frac{2-s_{\mathsf g}^2}{s_\sfv },\label{eq-def-cb-1}
\end{gather}
to simplify equations.
\end{subequations}
These expressions are already written in the unphysical frame with $\sigma=s^2_\sfv$.
The trace-reversed metric perturbation of $g'^{(\sfv)}_{\mu\nu}$ is given by $\bar h'_{\bar t\bar t}=\bar h'_{jk}= 0$, and 
 \begin{equation}
  \begin{split}
  \bar h'_{\bar tj}=\sum_{l=0}\pd_{jL}\frac{\check B_L}{r}-\sum_{l=1}&\pd_{L-1}\bigg( \frac{\check B_{jL-1}^{[2]}}{r} \\
  &-\epsilon_{jpq}\pd_{p}\frac{\check D_{qL-1}}{r}\bigg).
  \end{split}
\end{equation}
The first term on the right-hand side would make the computation very tedious, so  it would be better to simplify the metric components by performing the gauge transformation described by Eq.~\eqref{eq-sim}.
Then, one knows that the following good gauge
\begin{equation}
  \sW_L=\check B_L,\quad \sV_L=\sT_L=\sR_L=0.
\end{equation}
gives 
\begin{subequations}
  \label{eq-hbar-vv-f}
\begin{gather}
  \bar h''_{\bar t\bar t}= \sum_{l=0}\pd_L\frac{\check B_L^{[1]}}{r},\\
  \bar h''_{\bar tj}=-\sum_{l=1} \pd_{L-1}\left( \frac{\check B_{jL-1}^{[2]}}{r} -\epsilon_{jpq}\pd_{p}\frac{\check D_{qL-1}}{r}\right),\\
  \bar h''_{jk}= \delta_{jk}\sum_{l=0}\pd_L\frac{\check B_L^{[1]}}{r}.
\end{gather}
The \ae ther perturbation is 
\begin{gather}
  \mfv''^{\bar t}=-\sum_{l=0}\pd_L\frac{\check B_L^{[1]}}{r},\\
  \begin{split}
  \mfv''^j=\sum_{l=0}\pd_{jL}\frac{\mathring B_L}{r}-\sum_{l=1}&\pd_{L-1}\Bigg(\frac{\mathring B_{jL-1}^{[2]}}{r}\\
  &-\epsilon_{jpq}\pd_{p}\frac{\mathring D_{qL-1}}{r}\Bigg).
  \end{split}
\end{gather}
\end{subequations}
It is apparent that  $\bar h''_{\mu\nu}$ takes a very different form from the one in GR \cite{Blanchet:2020ngx}.
In particular, $\bar h''_{jk}$ is diagonal, while the corresponding perturbation in GR is tracefree \cite{Thorne:1980ru,Blanchet:2020ngx}.
These differences lead to the difficulty in determining a well-defined Newman-Unti coordinate system for this mode.

\subsection{Constructing pseudo-Newman-Unti coordinates}

In this subsection, one will determine the pseudo-Newman-Unti coordinates using the general scheme described in Section~\ref{sec-sch}.

As in the case of the tensor mode, let us first consider $\chi^\mu_{(\sfv)}$.
In Section~\ref{sec-sch}, $\chi^\mu_{(\sfv)}$ has been determined up to two sets of unspecified functions, $Z^a$ and $\sfR_{\{0\}}$, as the boundary conditions \eqref{eq-bdyc} were chosen to be relatively weak. 
Now, for the case of the vector mode, it is impossible to strengthen the condition \eqref{eq-bdyc-1-3}.
This is due to the form of $\bar h''_{\mu\nu}$ given by Eq.~\eqref{eq-hbar-vv-f}.
In particular, it is the following combination of the components that prevents the condition \eqref{eq-bdyc-1-3} from being tightened,
\begin{equation}
  \label{eq-gra-vec}
  \begin{split}
  e_a^j\left(\bar h''_{\bar tj}+\bar h''_{jk}n_k\right)=&\frac{e_a^j}{r}\sum_{l=1}(-1)^l n_{L-1}\left( \check B_{jL-1}^{[l+1]}\right.\\
  &\left.+\epsilon_{jpq}n_{p}\check D_{qL-1}^{[l]} \right)+\order{\frac{1}{r^2}}.
  \end{split}
\end{equation}
The left-hand side of the above expression appears in Eq.~\eqref{eq-tdg-ra}, which thus acquires a contribution at $\order{r^0}$.
One can check that $Z_a$ also appears at $\order{r^0}$ on the right-hand of Eq.~\eqref{eq-tdg-ra}.
As the leading order term of Eq.~\eqref{eq-gra-vec} is generally nonvanishing and time-dependent, one cannot choose a suitable $Z_a(\theta)$ to remove it, so $\tdg'^{(\sfv)}_{ra}$ has to start at $\order{r^0}$.
This also means that one shall leave $Z_a$ free, as a nonzero $Z_a$ does not violate the condition \eqref{eq-bdyc-1-3}.

Now, let us discuss the form of $\sfR_{\{0\}}$. 
As in the case of the tensor mode, one makes use of the vanishing of $\det(\tdg'^{(\sfv)}_{ab})$ at $\order{\tdr^3}$ to get
\begin{equation}
  \label{eq-r0-v}
  \sfR_{\{0\}}=\frac{1}{2}\left(\tdcd_aZ^a-\tdcd^2f\right).
\end{equation}
This equation takes the same form as Eq.~\eqref{eq-def-r0-ten} in the tensor sector, which guarantees the uniform treatment.
Therefore, $\chi_{(\sfv)}^\mu$ is completely fixed for the vector mode.
Just like $\chi_{(\sfg)}^\mu$ for the tensor mode, $\chi_{(\sfv)}^\mu$ is also parameterized by $T$, $Y^a$ and $Z^a$, so the asymptotic symmetry in the vector sector also includes the $\ebms$ and the subleading $\ebms$ symmetries.

The coordinate transformation from the Lorentz coordinates to the pseudo-Newman-Unti coordinates are given by
\begin{subequations}
  \label{eq-ctf-vec}
  \begin{gather}
    \sfU=\chi_{(\sfv)}^u+2\sum_{l=1}(-1)^l\sum_{k=1}^l\frac{la_{kl}}{k(l+k)}\frac{n_L\check B_L^{[l-k+1]}}{r^k},\\
    \begin{split}
    \sfR=\chi_{(\sfv)}^r+&\sum_{l=1}(-1)^l\sum_{k=1}^l\frac{l(l-k)+(k+1)^2}{(k+1)^2}\\
    &\times a_{kl}\frac{n_L\check B_L^{[l-k+1]}}{r^k},
    \end{split}\\
    \begin{split}
    \Theta^a=&\chi^a_{(\sfv)}
    -e_a^j\sum_{l=1}(-1)^l\sum_{k=1}^l\frac{a_{kl}}{(k+1)r^{k+1}}n_{L-1}\\
    &\times \left[ \frac{2l-k}{k}\check B_{jL-1}^{[l-k+1]}+\epsilon_{jpq}n_{p}\check D_{qL-1}^{[l-k]} \right].
    \end{split}
  \end{gather}
\end{subequations}
The leading order part of the coordinate transformation is given by $\chi_{(\sfv)}^\mu$, which describes the asymptotic symmetry, as elucidated in the previous paragraph.
The remaining parts of the transformation \eqref{eq-ctf-vec} are certainly different from Eq.~(3.2) in Ref.~\cite{Blanchet:2020ngx}.
They are given by functions of the multipole moments $\check B_L$ and $\check D_L$ related to the vector degrees of freedom in Einstein-\ae ther theory.

Substituting Eq.~\eqref{eq-ctf-vec} into Eq.~\eqref{eq-gc-pnu}, one can calculate the following metric components,
\begin{subequations}
  \label{eq-tdg-vec}
  \begin{gather}
    \tdg'^{(\sfv)}_{rr}=\frac{2}{\tdr} \bigg( \check B^{[1]}+2\sum_{l=1}(-1)^l\tdn_L\check B_L^{[l+1]} \bigg),\\
    \tdg'^{(\sfv)}_{ur}=-1+\frac{2\check B^{[1]}}{\tdr}+\frac{2}{\tdr} \sum_{l=1}(-1)^l\frac{l^2+3}{2}\tdn_L\check B_L^{[l+1]},\\
    \begin{split}
    \tdg'^{(\sfv)}_{ra}\equiv\Gamma_a=&-Z_a
    +\tilde e_a^j\sum_{l=1}(-1)^l \tdn_{L-1}\bigg(\check B_{jL-1}^{[l+1]}\\
    &+\tilde\epsilon_{jpq}\tdn_p\check D_{qL-1}^{[l]}  \bigg),\label{eq-tdg-ra-v}
    \end{split}
  \end{gather}
  where the symbol $\Gamma_a$ is defined for the upcoming discussion on asymptotic symmetries and vector memories.
  So these components are different from Eq.~\eqref{eq-def-nu-up}.
  Again, $\tdr$ is not null at a finite value, but $\tdg'^{(\sfv)}_{rr}$ approaches zero at the infinity.
  All of these components are even time dependent in general, in contrast with the corresponding terms in the tensor sector, and those in Newman-Unti gauge in GR.
  The Newman-Unti gauge can be imposed in the special case when $\check B_L^{[l+1]}=\check D_L^{[l]}=0$ under the gauge condition $Z_a=0$.
  As discussed below, this means that the presence of the vector GW prevents the construction of the Newman-Unti coordinates.
  The remaining metric components are
  \begin{gather}
    \tdg'^{(\sfv)}_{uu}=-1-\frac{1}{2}\tdcd_a(Y^a+\tdcd^2Y^a)
    +\order{\frac{1}{\tdr}},\\
    \begin{split}
    \tdg'^{(\sfv)}_{ua}=&-\tdcd_a \left( f+\frac{1}{2}\tdcd^2f+\frac{1}{2}\tdcd_bZ^b \right)\\
    &+\tilde e_a^j\sum_{l=1}(-1)^l\tdn_{L-1}\left\{ \bigg[ 1+\frac{1}{4}\right.\\
    &\times l(l+1)(2l-1) \bigg]\check B_{jL-1}^{[l+1]}\\
    &\left.+\frac{l^2+l+4}{4}\tilde\epsilon_{jpq}\tdn_p\check D_{qL-1}^{[l]}  \right\}
    +\order{\frac{1}{\tdr}},
    \end{split}\\
    \begin{split}
      \tdg'^{(\sfv)}_{ab}=&\tdr^2(\tdgamma_{ab}+2\tdcd_{\langle a}Y_{b\rangle})\\
      &+2\tdr(\tdcd_{\langle a}Z_{b\rangle}-\tdcd_{\langle a}\tdcd_{b\rangle}f)+\order{\tdr^0}.
    \end{split}
  \end{gather}
\end{subequations}
  The complete expressions are very tedious, and not very illuminating.
  So here, only the leading order terms are presented.
  The higher order terms in $1/\tdr$ are all expressed in terms of $\check B_L$, $\check D_L$, $\mathring B_L$, or $\mathring D_L$, defined in Eqs.~\eqref{eq-def-cb} and \eqref{eq-def-cb-1}.

  The terms containing $f$, $Y^a$, and $Z^a$ in Eq.~\eqref{eq-tdg-vec} tell us how the corresponding metric components transform under $\ebms$ and the subleading $\ebms$ transformations.
  The particularly interesting one for the discussion on memory effects is $\tdg'^{(\sfv)}_{ra}\equiv\Gamma_a$. 
  If one performs an infinitesimal subleading $\ebms$ transformation given by $\Theta^a\rightarrow\Theta^a-z_a/\tdr$ \footnote{Under this transformation, $Z_a\rightarrow Z_a+z_a$, referring to Eq.~\eqref{eq-def-chi-a}.}, then, 
  \begin{equation}
    \label{eq-sig-tf}
    \Gamma _a\rightarrow\Gamma _a-z_a,
  \end{equation}
  ignoring higher order terms in the field perturbations.
  So $\Gamma_a$ transforms nonlinearly, like $c_{ab}$ in the tensor sector.
  Note that $\Gamma_a$, ignoring the $Z_a$ part, is proportional to the angular component of $\Sigma_a$ at $\order{1/\tdr}$ by Eq.~\eqref{eq-sig-f}.
  That is, although the \ae{}ther field has no internal symmetries, its transformation under the diffeomorphism can be nontrivial.

  Finally, in the pseudo-Newman-Unti coordinates, the unphysical \ae ther field is
\begin{subequations}
  \begin{gather}
    \begin{split}
    \tilde\mfu'^u_{(\sfv)}=&1-\frac{\tdcd_aY^a}{2}+\frac{\mathring B^{[1]}}{\tdr}\\
    &+\frac{1}{\tdr}\sum_{l=0}(-1)^l(l^2-1)\tdn_L\check B_L^{[l+1]}+\order{\frac{1}{\tdr^2}},
    \end{split}\\
    \begin{split}
      \tilde\mfu'^r_{(\sfv)}=&-\frac{\mathscr D^2\mathscr D_aY^a}{4}-\frac{\mathring B^{[1]}}{\tdr}
      +\sum_{l=1}(-1)^l \\
      &\times\left[ 1+\frac{l(l-1)}{4} \right]
      \frac{l(l+1)}{2}\frac{\tdn_L\check B_L^{[l+1]}}{\tdr} \\
      &+\order{\frac{1}{\tdr^2}},
    \end{split}\\
    \begin{split}
      \tilde\mfu'^a_{(\sfv)}=&\frac{\mathscr D^a\mathscr D_bY^b}{2r}
      +\frac{\tilde e^a_j}{\tdr^2}  \sum_{l=1}(-1)^l\tdn_{L-1}
      \bigg\{\bigg( \mathring B_{jL-1}^{[l+1]}\\
      &+\tilde\epsilon_{jpq}\tdn_p\mathring D_{qL-1}^{[l]} \bigg) -\frac{1}{2} \left[ (2l-1)\check B_{jL-1}^{[l+1]}\right.\\
      &\left.+\tilde\epsilon_{jpq}n_{p}\check D_{qL-1}^{[l]} \right]\bigg\}+\order{\frac{1}{\tdr^3}}.
    \end{split}
  \end{gather}
\end{subequations}
We present the expressions for the components of the \ae ther field for the completeness.

\subsection{Vector memory effects}
\label{sec-vec-mm}

Now, let us calculate the geodesic deviation equation \eqref{eq-gdv-pnu} to study the memory effects excited by the vector mode.
It turns out that
\begin{equation}
  \label{eq-gdv-pnu-v}
  \frac{\ud^2\mathsf S}{\ud\tau^2}=-s_\sfv^2\tilde R_{uru\hat a}S^{\hat a},\quad\frac{\ud^2S_{\hat a}}{\ud\tau^2}=-s_\sfv^2\tilde R_{uru\hat a}\mathsf S,
\end{equation}
as $\tilde R_{urur}=\tilde R_{u\hat au\hat b}=0$ at $\order{1/\tdr}$, and
\begin{equation}
  \label{eq-rie-v}
  \begin{split}
  \tilde R_{uru\hat a}=&-\frac{c_+s_\sfg^2}{2s_\sfv\tdr}\tilde e_{\hat a}^j\sum_{l=1}(-1)^l\tdn_{L-1} \bigg( B_{jL-1}^{[l+3]}\\
  &+\tilde\epsilon_{jpq}\tdn_pD_{qL-1}^{[l+2]} \bigg)+\order{\frac{1}{\tdr^2}}.
  \end{split}
\end{equation}
So there is a mixing between the longitudinal mode $\mathsf S$ and the transverse modes $S^a$.
This is just the special feature of the vector polarizations.
Since when $\check B_L^{[l+1]}=\check D_L^{[l]}=0$, it is permissible to take the Newman-Unti gauge, the above discussion implies that in that case, there are no vector GWs at $\tdr\rightarrow\infty$, noting that $\check B_L\propto B_L$ and $\check D_L\propto D_L$ by Eqs.~\eqref{eq-def-cb} and \eqref{eq-def-cb-1}.
Therefore, in some sense, it is the presence of the vector GWs that impedes the use of the Newman-Unti coordinates in the vector sector.

Comparing Eqs.~\eqref{eq-rie-v}, \eqref{eq-sig-f}, and \eqref{eq-tdg-ra-v}, one can rewrite 
\begin{equation}
  \label{eq-rie-v2}
  \tilde R_{uru\hat a}=-\frac{c_+s_\sfg^2}{2s_\sfv\sfk_1}\Gamma _{\hat a}^{[2]},
\end{equation}
at $\order{1/\tdr}$, with $\sfk_1$ defined in Eq.~\eqref{eq-def-cb}.
This is expected, as the vector polarizations are excited by the vector modes $\Sigma_j$, and one can show that $\tde_a^j\Sigma_j=\Gamma_a/\sfk_1\tdr+\order{1/\tdr^2}$.
So integrating Eq.~\eqref{eq-gdv-pnu-v} twice results in
\begin{equation}
  \label{eq-vmm-l}
    \begin{split}
    \Delta \mathsf S=&\left.\frac{\ud\mathsf S}{\ud\tau}\right|_0\Delta\tdu+\frac{c_+s_\sfv s_\sfg^2}{2\sfk_1}\Delta\Gamma _{\hat a} S^{\hat a}|_0\\
    &+\frac{c_+s_\sfv s_\sfg^2}{2\sfk_1} \Bigg\{  \left[ \Gamma _{\hat a}(\tdu_f)+\Gamma _{\hat a}(\tdu_0) \right]\Delta\tdu\\
    &-2\int_{\tdu_0}^{\tdu_f}\Gamma _{\hat a}\ud\tdu  \Bigg\}\left.\frac{\ud S^{\hat a}}{\ud\tau}\right|_0,
    \end{split}
\end{equation}
and an exactly similar expression for $\Delta S_{\hat a}$ with $\mathsf S$ and $S^{\hat a}$ exchanged in the above equation.
Note that one should evaluate this equation up to $1/\tdr$.
Here, Eq.~\eqref{eq-vmm-l} is written in terms of $\Gamma_a$ so that it is easy to associate the memory effect with the symmetry (see Eq.~\eqref{eq-sig-tf}).
This equation is similar to Eq.~\eqref{eq-def-tmm} for the tensor mode, so one could call the term proportional to $\Delta\Gamma_{\hat a}$ the (leading) displacement memory effect of the vector mode, and the one proportional to the integral of $\Gamma_{\hat a}$ the subleading displacement memory effect.
By the transformation law \eqref{eq-sig-tf}, one may state that the (leading) displacement memory of the vector mode is associated with the subleading transformation,
\begin{equation}
  \label{eq-vec-dis}
  \Delta\Gamma _a=-z_a,
\end{equation}
similar to Eq.~\eqref{eq-ten-dis} for the tensor displacement memory effect.
So the (leading) vector displacement memory effect may be interpreted as the vacuum transition $\Gamma _a(u_0)\rightarrow\Gamma _a(u_f)$ in the vector sector.
This observation serves as one reason for keeping $Z_a$ in $\chi^\mu_{(\sfm)}$.
It seems that the subleading vector displacement memory $\int_{\tdu_0}^{\tdu_f}\Gamma _a\ud \tdu$ has no close relation with the asymptotic symmetry.
    
Since any vector field on a unit 2-sphere can be decomposed into its electric and magnetic parts, e.g.,
\begin{equation}
  \label{eq-f-dec}
  \Gamma_a=\tdcd_a\mu+\tilde\epsilon_{ab}\tdcd^b\nu,
\end{equation}
one may call $\Delta\mu$ the electric-type vector memory effect, and $\Delta\nu$ the magnetic-type.
Due to Eq.~\eqref{eq-vec-dis}, one writes,
\begin{equation}
  \label{eq-e-m-z}
  \Delta\mu=-\mfx,\quad\Delta\nu=-\mfy,
\end{equation}
where one decomposes $z_a=\tdcd_a\mfx+\tilde\epsilon_{ab}\tdcd^b\mfy$.
This means that the electric-type vector memory effect can be viewed as the vacuum transition caused by the electric parity part of the subleading $\ebms$ symmetry $z_a=\tdcd_a\mfx$, while the magnetic-type memory is the vacuum transition associated with the magnetic parity part of $z_a=\tilde\epsilon_{ab}\tdcd^b\mfy$. 
If one expands $\Delta\mu$ and $\Delta\nu$ in the similar manner to $\Delta\sfc$ in Eq.~\eqref{eq-def-Dcexp},  calculation shows that 
\begin{gather}
\Delta\mu_L=\frac{(-1)^l}{l}\Delta\check B_L^{[l+1]},\\
\Delta\nu_L=\frac{(-1)^l}{l}\Delta\check D_L^{[l+1]},
\end{gather}
for $l\ge1$, according to the definition of $\Gamma_a$.
As a comparison, the velocity kick memory effect in Maxwell's electrodynamics is related to the change in the electric part of the leading order term of the radiative mode, since the gauge symmetry of the 4-potential $\mathscr A_\mu\rightarrow\mathscr A_\mu+\partial_\mu\vartheta$ involves the electric part of $\mathscr A_\mu$ \cite{Strominger:2018inf}.
Here, the linear analysis shows that the electric-type and magnetic-type memories both exist in a generic process in Einstein-\ae{}ther theory.

Similarly, one can decompose the subleading displacement memory effect $\int_{u_0}^{u_f}\Gamma_a\ud\tdu$ into its electric ($\int_{\tdu_0}^{\tdu_f}\mu\ud\tdu$) and magnetic ($\int_{\tdu_0}^{\tdu_f}\nu\ud\tdu$) parts.
Their relations with the symmetries are trivial at least in the linearized case.

Finally, in the linearized theory, one could not get the constraint equations for the vector memory effects, either.

\section{Pseudo-Newman-Unti coordinates for scalar mode}
\label{sec-pnu-s}

In this section, the pseudo-Newman-Unti coordinates will be obtained for the scalar mode following the general scheme in Section~\ref{sec-sch}.
The structure is again similar to Section~\ref{sec-pnu-t} for the tensor mode.

\subsection{Vacuum multipolar solution}

For the scalar mode, one can simply set, for example, 
\begin{equation}
  \label{eq-def-dec-s}
  \Omega=\sum_{l=0}\pd_L\frac{A_L(s_{\mathsf s}t-r)}{r}.
\end{equation}
This solves Eq.~\eqref{eomscl2}.
With Eqs.~\eqref{eq-s-1} and \eqref{eq-s-2}, one has
\begin{gather}
  \Phi=\frac{c_{14}-2c_+}{2-c_{14}}\sum_{l=0}\pd_L\frac{\dot A_L}{r},\\
  \Psi=\frac{2c_{14}(c_+-1)}{2-c_{14}}\sum_{l=0}\pd_L\frac{\dot A_L}{r}.
\end{gather}
There are no constraints on $A_L$'s from the equation of motion \eqref{eomscl2}.

\subsection{Gauge fixing}

Similarly to the tensor and vector modes, one wants to fix the gauge before obtaining the pseudo-Newman-Unti coordinates.
Again, one may set $\varepsilon_j=0$ and $\omega=\gamma=0$, and perform the coordinate transformation 
$t\rightarrow \bar t=s_\sfs t,\,x^j\rightarrow x^j$.
In this way, one knows that 
\begin{subequations}
\begin{gather}
  g'^{(\sfm)}_{\bar t\bar t}=-1-\frac{4}{2-c_{14}}\frac{s_\sfs ^2}{s_{\mathsf g}^2}\sum_{l=0}\pd_L\frac{\mathcal A_L^{[2]}}{r},\\
  g'^{(\sfm)}_{\bar tj}=0,\\
  g'^{(\sfm)}_{jk}
  =\delta_{jk} -\frac{2c_{14}}{2-c_{14}}\frac{s_\sfs ^2}{s_{\mathsf g}^2}\sum_{l=0}\pd_L\frac{\mathcal A_L^{[2]}}{r}\delta_{jk}+2\sum_{l=0}\pd_{jkL}\frac{\mathcal A_L}{r},
\end{gather}
and 
\begin{gather}
  \mfu'^{\bar t}_{(\sfm)}=1-\frac{2}{2-c_{14}}\frac{s_\sfs ^2}{s_{\mathsf g}^2}\sum_{l=0}\pd_L\frac{\mathcal A_L^{[2]}}{r},\\
  \mfu'^j_{(\sfm)}=0.
\end{gather}
\end{subequations}
In the above expressions, the superscript $[n]$ means to take the $n$-th order derivative with respect to $\bar t$, and $\mathcal A_L=\int A_L\ud t$. 
These expressions are already written in the unphysical frame with $\sigma=s_\sfs^2$.
The trace-reversed metric perturbation can be read off, and given by $\bar h'_{\bar tj}=0$, and
\begin{subequations}
  \label{eq-hbar-s}
 \begin{gather}
  \bar h'_{\bar t\bar t}= \left( 1-\frac{2+3c_{14}}{2-c_{14}}\frac{s_\sfs ^2}{s_{\mathsf g}^2} \right)\sum_{l=0}\pd_L\frac{\mathcal A_L^{[2]}}{r},\\
  \bar h'_{jk}=-\delta_{jk}   \left(1+\frac{s_\sfs ^2}{s_{\mathsf g}^2}\right) \sum_{l=0}\pd_L\frac{\mathcal{A}_L^{[2]}}{r} +2\sum_{l=2}\pd_{jkL}\frac{\mathcal A_L}{r}.
\end{gather}   
\end{subequations}
A further gauge transformation given by Eq.~\eqref{eq-def-inf-gtf} should now depend on  $s_{\mathsf s}t-r$, and help simplify the following calculation.
Then, the good gauge in the scalar sector is
\begin{equation}
  \sV_L=\mathcal A_L,\quad \sW_L=\sV_L^{[1]}=\mathcal A_L^{[1]}.
\end{equation}
After this transformation, the trace-reversed metric perturbation transforms to 
\begin{subequations}
  \label{eq-hbar-ss-f}
\begin{gather}
  \bar h''_{\bar t\bar t}= (2\mathscr F_1-\mathscr F_2)\sum_{l=0}\pd_L\frac{\mathcal A_L^{[2]}}{r},\\
  \bar h''_{\bar tj}=0,\\
  \bar h''_{jk}= \delta_{jk}  \mathscr F_2\sum_{l=0}\pd_L\frac{\mathcal A_L^{[2]}}{r},
\end{gather}
where
  \begin{equation}
    \mathscr F_1=1-\frac{2+c_{14}}{2-c_{14}}\frac{s_{\mathsf s}^2}{s_{\mathsf g}^2},\quad \mathscr F_2=1-\frac{s_{\mathsf s}^2}{s_{\mathsf g}^2}.
  \end{equation}
The \ae ther perturbation is given by 
\begin{gather}
  \mfv''^{\bar t}=\frac{\mathscr F_1+\mathscr F_2-4}{2}\sum_{l=0}\pd_L\frac{\cA_L^{[2]}}{r},\\
  \mfv''^j=\sum_{l=0}\pd_{jL}\frac{\cA_L^{[1]}}{r}.
\end{gather}
\end{subequations}
Again, $\bar h''_{\mu\nu}$ differs from the one in GR \cite{Thorne:1980ru,Blanchet:2020ngx} in several aspects.
Here, $\bar h''_{jk}$ is also diagonal as for the case of the vector mode, and $\bar h''_{\bar tj}=0$.
Finally, there are no constant $\cA_L$'s, while in GR, $M$, $P_j=M_j^{[1]}$, and $S_j$ are constant.

\subsection{Constructing pseudo-Newman-Unti coordinates}

Finally, let us determine the pseudo-Newman-Unti coordinates for the scalar mode. 
Again, we start with $\chi_{(\sfs)}^\mu$, i.e., we want to check if $Z^a$ and $\sfR_{\{0\}}$ can be fixed.
As one can check, $Z_a$ appears in Eq.~\eqref{eq-tdg-ra}, and one can set it free to make the condition \eqref{eq-bdyc-1-3} be satisfied.
Now, to determine $\sfR_{\{0\}}$, one uses the condition \eqref{eq-bdyc-3}, and thus one should calculate Eq.~\eqref{eq-tdg-ab-f}.
It turns out that
\begin{equation*}
  \label{eq-def-r0-s-0}
  \sfR_{\{0\}}=\frac{1}{2}\left(\tdcd_aZ^a-\tdcd^2f\right)+\frac{\mathscr F_\Delta}{2}\sum_{l=0}(-1)^l\tdn_L\cA_L^{[l+2]},
\end{equation*}
  where 
  \begin{equation}
    \mathscr F_\Delta=\mathscr F_1-\mathscr F_2.
  \end{equation}
However, this expression does not take the similar forms as Eqs.~\eqref{eq-def-r0-ten} and \eqref{eq-r0-v} for the tensor and vector modes, respectively.
Moreover, $\sfR_{\{0\}}$ depends on the scalar perturbation.
Thus, it is better to redefine $\sfR_{\{0\}}$ for the scalar mode by moving the scalar dependent part to $\sfR'$ introduced in Eq.~\eqref{eq-def-exp-r} so that
\begin{equation}
  \label{eq-def-r0-s}
  \sfR_{\{0\}}=\frac{1}{2}\left(\tdcd_aZ^a-\tdcd^2f\right).
\end{equation}
In this way, $\sfR_{\{0\}}$ is completely independent of the field perturbation, and truly becomes one component of  the generator of an asymptotic symmetry, which shall not be a function of the field perturbations.
$\sfR_{\{0\}}$ now also agrees with Eqs.~\eqref{eq-def-r0-ten} and \eqref{eq-r0-v}, confirming the uniform treatment of the different propagating modes.
Therefore,  $\chi^\mu_{(\sfs)}$ is again described by $T$, $Y^a$ and $Z^a$. 
The asymptotic symmetry in the scalar sector also includes the $\ebms$ symmetry and the subleading $\ebms$ symmetry.

Now, the complete coordinate transformation is given by
\begin{subequations}
  \label{eq-ctf-s}
  \begin{gather}
    \sfU=\chi^u_{(\sfs)}+\mathscr F_1\sum_{l=1}(-1)^l\sum_{k=1}^l\frac{a_{kl}}{k}\frac{n_L\cA_L^{[l-k+2]}}{r^k},\\
    \begin{split}
    \sfR=&\chi^r_{(\sfs)}+\sum_{l=1}(-1)^l\sum_{k=1}^l \bigg[ \mathscr F_2+\frac{l-k}{2}\\
    &\times\frac{l+k+1}{(k+1)^2}\mathscr F_1 \bigg]
    \frac{a_{kl}}{k}\frac{n_L\cA_L^{[l-k+2]}}{r^k},
    \end{split}\\
    \begin{split}
    \Theta^a=&\chi^a_{(\sfs)}-\frac{e^a_j}{r}\sum_{l=1}(-1)^l\\
    &\times\sum_{k=1}^l\frac{la_{kl}}{k(k+1)}\frac{n_L\cA_L^{[l-k+2]}}{r^k},
    \end{split}
  \end{gather}
\end{subequations}
based on the general scheme presented in Section~\ref{sec-sch}.
Under the coordinate transformation \eqref{eq-ctf-s}, one gets,
\begin{subequations}
  \begin{gather}
    \tdg'^{(\sfs)}_{rr}=2\mathscr F_1\sum_{l=0}(-1)^l\frac{\tdn_L\cA_L^{[l+2]}}{\tdr},\\
    \tdg'^{(\sfs)}_{ra}=-Z_a,\\
    \begin{split}
    \tdg'^{(\sfs)}_{ur}=-1+&\sum_{l=0}(-1)^l \left( \mathscr F_2+\frac{l^2+l+2}{2}\mathscr F_1 \right)\\
    &\times\frac{\tdn_L\cA_L^{[l+2]}}{\tdr},
    \end{split}
  \end{gather}
so, at the finite $\tdr$, $\tpd_r$ is not a null vector.
In the special solution with $\cA_L^{[l+2]}=0$ under the gauge condition $Z_a=0$, the Newman-Unti gauge is recovered.
The remaining metric components have the following leading order terms,
  \begin{gather}
    \tdg'^{(\sfs)}_{uu}=-1-\frac{1}{2}\tdcd_a(Y^a+\tdcd^2Y^a)+\order{\frac{1}{\tdr}},\\
    \tdg'^{(\sfs)}_{ua}=-\tdcd_a \left( f+\frac{1}{2}\tdcd^2f+\frac{1}{2}\tdcd_bZ^b \right)+\order{\frac{1}{\tdr}},\\
    \begin{split}
      \label{eq-tdg-ab-s}
      \tdg'^{(\sfs)}_{ab}=&\tdr^2(\tdgamma_{ab}+2\tdcd_{\langle a}Y_{b\rangle})+\tdr\Bigg[2\tdcd_{\langle a}Z_{b\rangle}\\
      &-2\tdcd_{\langle a}\tdcd_{b\rangle}f+\tdgamma_{ab} \mathscr F_\Delta\sum_{l=0}(-1)^l\tdn_L\cA_L^{[l+2]}\Bigg]\\
      &+\order{\tdr^0},
    \end{split}
  \end{gather}
\end{subequations}
From these equations, one knows that the coefficients of the higher order terms in $1/\tdr$ transform trivially under the $\ebms$ and subleading $\ebms$ transformations in the linearized theory.
In Eq.~\eqref{eq-tdg-ab-s}, the term proportional to $\mathscr F_\Delta$ may seem transform nontrivially. 
But since this term is a trace, while the remaining ones in the squared brackets are tracefree, then it truly remains invariant under the $\ebms$ and the subleading $\ebms$ symmetries.

Finally, the unphysical \ae ther field is
\begin{subequations}
  \begin{gather}
    \begin{split}
    \tilde\mfu'^u=&1-\frac{\tdcd_aY^a}{2}+\frac{1}{2\tdr}\sum_{l=0}(-1)^l[(l^2+l+1)\mathscr F_1\\
    &+\mathscr F_2-2]\tdn_L\cA_L^{[l+2]}+\order{\frac{1}{\tdr^2}},\\
      \tilde \mfu'^r=&-\frac{\tdcd^2\tdcd_aY^a}{4}+\order{\frac{1}{\tdr}},
    \end{split}\\
      \begin{split}
      \tilde\mfu'^a=&\frac{\tdcd^a\tdcd_bY^b}{2\tdr}\\
      &-\mathscr F_1\tilde e^a_j\sum_{l=1}(-1)^l\frac{l}{2}\frac{\tdn_L\cA_{jL-1}^{[l+2]}}{\tdr^{2}}+\order{\frac{1}{\tdr^2}}.
      \end{split}
  \end{gather}
\end{subequations}

\subsection{Scalar memory effects}
\label{sec-sca-mm}

After some mathematical manipulation, Eq.~\eqref{eq-gdv-pnu} becomes
  \begin{equation}
    \label{eq-gdv-pnu-s}
    \frac{\ud^2\mathsf S}{\ud\tau^2}=-s^2_\sfm\tilde R_{urur}\mathsf S,\quad
    \frac{\ud^2S_{\hat a}}{\ud\tau^2}=-s_\sfm^2\tilde R_{u\hat au\hat b}S^{\hat b},
  \end{equation}
at the linear order in $1/\tdr$ for the scalar mode, where
  \begin{gather*}
  \tilde R_{urur}=\mathscr F_3\sum_{l=0}(-1)^l\frac{\tdn_L\cA_L^{[l+4]}}{\tdr}+\order{\frac{1}{\tdr^2}},\\
  \tilde R_{uru\hat a}=\order{\frac{1}{\tdr^2}},\\
  \tilde R_{u\hat au\hat b}=\mathscr F_4\tdgamma_{ab}\sum_{l=0}(-1)^l\frac{\tdn_L\cA_L^{[l+4]}}{\tdr}+\order{\frac{1}{\tdr^2}},\\
  \end{gather*}
and
\begin{gather*}
  \mathscr F_3=\frac{1}{2-c_{14}} \left[ \left( 1+\frac{s_\sfs^2}{s_\sfg^2} \right)c_{14}-2c_+ \right],\\
  \mathscr F_4=\frac{c_{14}}{2-c_{14}}\frac{s_\sfs^2}{s_\sfg^2}.
\end{gather*}
So there is no mixing between the longitudinal mode $\mathsf S$ and the transverse modes $S^{\hat a}$.
As discussed above, when $\cA_L^{[l+2]}=0$, it is possible to impose Newman-Unti gauge condition.
In this case, the geodesic deviation equation is trivial, which means that the scalar GW is absent at $\tdr\rightarrow\infty$.
Therefore, the scalar GW prevents one from constructing the Newman-Unti coordinate system in the scalar sector.

One can also rewrite 
\begin{equation}
 \tilde R_{urur}=\mathscr F_3\Omega^{[3]},\quad \tilde R_{u\hat au\hat b}=\mathscr F_4\tdgamma_{ab}\Omega^{[3]},
\end{equation}
at the linear order in $1/\tdr$.
Indeed, the scalar mode excites the longitudinal ($\mathsf S$) and breathing ($S^a$) polarizations  \cite{Eardley:1974nw,Eardley:1973br,Will:2014kxa,Hou:2017bqj,Gong:2018cgj}.
Therefore, the integration  of Eq.~\eqref{eq-gdv-pnu-s} gives 
\begin{equation*}
  \label{eq-def-smm}
  \begin{split}
  \Delta\mathsf S=&\left.\mathsf S^{[1]}\right|_0\Delta\tdu-\mathscr F_3\Delta\Omega^{[1]}\mathsf S|_0\\
  &-\mathscr F_3 \left\{ [\Omega^{[1]}(\tdu_f)+\Omega^{[1]}(\tdu_0)]\Delta\tdu-2\Delta\Omega \right\}\left.\mathsf S^{[1]}\right|_0,
  \end{split}
\end{equation*}
with a similar equation for $\Delta S_{\hat a}$ with $\mathsf S$ and $\mathscr F_3$ replaced by $S_{\hat a}$ and $\mathscr F_4$, respectively.
Again, this equation shall be evaluated at the order of $1/\tdr$, as in the case of the tensor and vector modes.
One would like to call $\Delta\mathsf S$ and $\Delta S_{\hat a}$ the longitudinal and the breathing memory effects, with the term proportional to $\Delta\Omega^{[1]}$ the leading memory effects, and the one proportional to $\Delta\Omega$ the subleading effects.
Unlike the tensor or vector memory effects, there seems no clear relation between the asymptotic symmetries and the scalar memories.

\section{Discussion and conclusion}
\label{sec-con}

In this work, the asymptotic analysis of the vacuum Einstein-\ae ther theory has been done in the linear case.
Since there are three types of radiated modes, and they generally travel at different speeds, the asymptotic analysis has to be performed separately for each mode. 
These radiative modes satisfy the similar d'Alembertian equations with the speed of light replaced by their respective speeds.
This implies that one may analyze their solutions using a general scheme such that a suitable coordinate system, named the pseudo-Newman-Unti coordinates ($\tdu,\tdr,\tdth^a$), can be constructed for each mode.
In this coordinate system, the components of the metric and \ae ther fields can be written as series expansions in $1/\tdr$, without $\ln\tdr$-terms.
Although $\tdr$ is not always a null direction, it is approximately null, measured by a suitably defined unphysical metric $\tdg'^{(\sfm)}_{\mu\nu}$, as $\tdr\rightarrow\infty$.
It turns out that, at the leading orders in $1/\tdr$, the asymptotic symmetry is parameterized by three sets of functions of angles, $T$, $Y^a$, and $Z^a$, for all modes.
As in GR, BD, and dCS, $T$ generates the supertranslation transformation, and $Y^a$ belongs to the Lorentz algebra or its extensions.
In addition, $Z^a(\tdth^b)$ parameterizes the transformation subleading relative to $Y^a$.
Therefore, the asymptotic symmetry group of Einstein-\ae ther theory contains the $\ebms$ and the subleading $\ebms$ symmetries for each radiative mode, generally larger than that in GR, BD, or dCS.

One may find it uneasy that the asymptotic symmetry group of Einstein-\ae ther theory is larger. 
So let us explain it here.
The asymptotic symmetry is actually the residual gauge symmetry, obtained once one fixes the asymptotic behaviors of the metric components.
Since all metric theories mentioned so far enjoy the diffeomorphism invariance, the stronger the boundary conditions of the metric are, the smaller the residual gauge symmetry group is.
As discussed in Section~\ref{sec-sch}, one could not impose the Newman-Unti gauge condition \eqref{eq-def-nu-up}, otherwise, there would be $\ln\tdr$ or even $\tdr\ln\tdr$ terms appearing in the metric and \ae ther fields for the vector and scalar modes, relative to their respective leading order terms.
These blowing up terms render the linearization inconsistent.
Therefore, in this work, one may impose weaker conditions \eqref{eq-bdyc}, and this leads to the larger asymptotic symmetry group of Einstein-\ae ther theory.

In the main text, it was implicitly assumed that three kinds of radiative modes travel at different speeds, so the construction of the pseudo-Newman-Unti coordinates was done separately for each mode.
Although we used the same symbols, e.g., $T$, $Y^a$, and $Z^a$, they actually have different meanings in sections~\ref{sec-pnu-t}, \ref{sec-pnu-v}, and \ref{sec-pnu-s}.
When all modes share the same velocity, not necessarily the physical speed of light, one can construct a single pseudo-Newman-Unti coordinate system.
Formally, one just adds up the corresponding equations.
For examples, one can add up the right-hand sides of Eqs.~\eqref{eq-ctf-ten}, \eqref{eq-ctf-vec}, and \eqref{eq-ctf-s} to get the coordinate transformation.
Of course, in doing so, one should treat all symmetric-tracefree tensors, $M_L$, $S_L$, $B_L$, $D_L$, $\cA_L$, etc., functions of $st-r$ with $s$ a common speed.
One should also drop the subscripts of $\chi^\mu_{(\sfg,\sfv,\sfs)}$, and formally, keep one copy of them \footnote{In other words, $3\chi^\mu$ still parameterizes the same asymptotic symmetry, and the factor 3 can be absorbed by a redefinition.}.
Therefore, the asymptotic symmetry group is still $\ebms$ and its subleading version.

The memory effects have also been analyzed, by integrating the geodesic deviation equation for each radiative mode. 
Like in GR, BD, and dCS, the tensor modes excite the displacement, spin and CM memories. 
The displacement memory effect is also closely related to the supertranslation, and can be treated as the vacuum transition in the tensor sector. 
As in the linear theory, one cannot determine the complete transformation rules of various quantities, in particular, the shear tensor $c_{ab}$, it is impossible to find the constraint equations on these memory effects, which involve terms quadratic in $c_{ab}$ and $N_{ab}$.
For the vector and scalar modes, there are also (leading) displacement memories. 
Like the tensor displacement memory effect, the (leading) vector displacement memory can also be associated with the subleading transformation generated by $Z^a$.
Both the electric-type and magnetic-type vector memory effects take place in a generic physical process.
The subleading displacement memories could also be defined using the integrated geodesic deviation equations.
The vector subleading memories can be decomposed into the electric and magnetic parts.
Finally, there seems to be no relation between the scalar memories and the asymptotic symmetries.
One may seek for the dual formalism similar to those described in Refs.~\cite{Campiglia:2017dpg,Campiglia:2018see,Seraj:2021qja}, which is beyond the scope of the current work.

There are several questions that would be answered in the future, in addition to the ones mentioned previously.
For example, $\chi^\mu_{(\sfm)}$ has been defined at the leading orders in $\tdr$, 
so whether it has higher order corrections like $\xi^\mu_\text{BMS}$ in GR \cite{Barnich:2010eb} is such a question. 
Once one obtains the complete expression for $\chi^\mu_{(\sfm)}$, one can work out the Lie algebra consistently.
This would help determine the Noether charges and flux-balance laws, using Hamiltonian formalism \cite{Wald:1999wa}.
Also, is there any soft theorem related to $Z^a$, and what are the conserved quantities associated with $\ebms$, and the flux-balance laws? 
All of these questions may involve the complete nonlinear analysis, which will be done in the future.

\begin{acknowledgements}
The authors were grateful for the discussion with Zoujian Cao, Alexander Grant, Jiang Long, Pujian Mao and Chao Zhang.
This work was supported by the National Natural Science Foundation of China under grant Nos.~11633001 and 11920101003, and the Strategic Priority Research Program of the Chinese Academy of Sciences, grant No.~XDB23000000.
S. H. was supported by the National Natural Science Foundation of China under Grant No.~12205222, and by the Fundamental Research Funds for the Central Universities under Grant No.~2042022kf1062.
\end{acknowledgements}

\bibliography{MemoryAether_separate_v8.bbl}

\end{document}